\documentclass[iop,revtex4]{emulateapj}

\usepackage{graphics}
\usepackage{epsfig}
\usepackage{natbib}
\usepackage{lscape}
\usepackage{enumerate}
\usepackage{hyperref}

\newcommand{\kms}{km~s\ensuremath{^{-1}}}
\newcommand{\msun}{$M_{\odot}$}
\newcommand{\nuvr}{NUV-$r$}
\newcommand{\mh}{$M_{H_2}$}
\newcommand{\mhi}{$M_{HI}$}
\newcommand{\mstar}{$M_{\ast}$}
\newcommand{\must}{$\mu_{\ast}$}

\newcommand{\rmol}{$R_{mol}$}
\newcommand{\tdep}{$t_{dep}({\rm H_2})$}
\newcommand{\tdepHI}{$t_{dep}({\rm HI})$}
\newcommand{\fgas}{$f_{H_2}$}
\newcommand{\fhi}{$f_{HI}$}
\newcommand{\xco}{$\alpha_{CO}$}
\newcommand{\ms}{SFR-M$_{\ast}$}
\newcommand{\deltams}{$\Delta$(MS)}

\newcommand{\hi}{{H{\sc i}}}
\newcommand{\about}{$\sim$}
\newcommand{\edit}{ }

\shorttitle{xCOLD GASS: an IRAM legacy}
\shortauthors{Saintonge et al.}

\begin{document}

\title{\lowercase{x}COLD GASS: the complete IRAM-30\lowercase{m} legacy survey of molecular gas \\ 
for galaxy evolution studies}

\author{Am\'{e}lie Saintonge\altaffilmark{1},  
Barbara Catinella\altaffilmark{2}, Linda J. Tacconi\altaffilmark{3}, Guinevere Kauffmann\altaffilmark{4}, 
 Reinhard Genzel\altaffilmark{3},  \\
 Luca Cortese\altaffilmark{2},  Romeel Dav\'e\altaffilmark{5,6,7}, Thomas J. Fletcher\altaffilmark{1}, Javier Graci\'{a}-Carpio\altaffilmark{3}, Carsten Kramer\altaffilmark{8},  \\
 Timothy M. Heckman\altaffilmark{9}, Steven Janowiecki\altaffilmark{2}, Katharina Lutz\altaffilmark{10}, David Rosario\altaffilmark{11}, David Schiminovich\altaffilmark{12}, \\ 
 Karl Schuster\altaffilmark{13}, Jing Wang\altaffilmark{14}, Stijn Wuyts\altaffilmark{15}, Sanchayeeta Borthakur\altaffilmark{16}, \\
 Isabella Lamperti\altaffilmark{1} \& Guido W. Roberts-Borsani\altaffilmark{1} }

\altaffiltext{1}{University College London, Gower Street, London WC1E 6BT, UK}
\altaffiltext{2}{ICRAR, M468, The University of Western Australia, 35 Stirling Highway, Crawley, Western Australia 6009, Australia}
\altaffiltext{3}{Max-Planck Institut f\"ur extraterrestrische Physik, 85741 Garching, Germany}
\altaffiltext{4}{Max-Planck Institut f\"ur Astrophysik, 85741 Garching, Germany}
\altaffiltext{5}{University of the Western Cape, Bellville, Cape Town 7535, South Africa}
\altaffiltext{6}{South African Astronomical Observatories, Observatory, Cape Town 7925, South Africa}
\altaffiltext{7}{Institute for Astronomy, Royal Observatory, Edinburgh EH9 3HJ, UK}
\altaffiltext{8}{Instituto Radioastronom\'{i}a Milim\'{e}trica, Av. Divina Pastora 7, Nucleo Central, 18012 Granada, Spain}
\altaffiltext{9}{Johns Hopkins University, Baltimore, Maryland 21218, USA}
\altaffiltext{10}{Centre for Astrophysics and Supercomputing, Swinburne University of Technology, Hawthorn, Victoria 3122, Australia}
\altaffiltext{11}{Centre for Extragalactic Astronomy, Department of Physics, Durham University, South Road, Durham DH1 3LE, UK}
\altaffiltext{12}{Department of Astronomy, Columbia University, New York, NY 10027, USA}
\altaffiltext{13}{Institut de Radioastronomie Millim\'{e}trique, 300 Rue de la piscine, 38406 St Martin d'H\`{e}res, France}
\altaffiltext{14}{Kavli Institute for Astronomy and Astrophysics, Peking University, Beijing 100871, China}
\altaffiltext{15}{Department of Physics, University of Bath, Claverton Down, Bath, BA2 7AY, UK}
\altaffiltext{16}{School of Earth and Space Exploration, Arizona State University, Tempe, AZ 85287, USA }

\begin{abstract}
We introduce xCOLD GASS, a legacy survey providing a census of molecular gas in the local Universe. Building upon the original COLD GASS survey, we present here the full sample of 532 galaxies with CO(1-0) measurements from the IRAM-30m telescope. The sample is mass-selected in the redshift interval $0.01<z<0.05$ from SDSS, and therefore representative of the local galaxy population with \mstar$>10^9$\msun. The CO(1-0) flux measurements are complemented by observations of the CO(2-1) line with both the IRAM-30m and APEX telescopes, HI observations from Arecibo, and photometry from SDSS, WISE and GALEX. Combining the IRAM and APEX data, we find that the CO(2-1) to CO(1-0) luminosity ratio for integrated measurements is $r_{21}=0.79\pm0.03$, with no systematic variations across the sample. The CO(1-0) luminosity function is constructed and best fit with a Schechter function with parameters {$L_{\mathrm{CO}}^* = (7.77\pm2.11) \times 10^9\,\mathrm{K\,km\,s^{-1}\, pc^{2}}$, $\phi^{*} = (9.84\pm5.41) \times 10^{-4} \, \mathrm{Mpc^{-3}}$ and $\alpha = -1.19\pm0.05$}. With the sample now complete down to stellar masses of $10^9$\msun, we are able to extend our study of gas scaling relations and confirm that both molecular gas fractions (\fgas) and depletion timescale (\tdep) vary with specific star formation rate (or offset from the star-formation main sequence) much more strongly than they depend on stellar mass. Comparing the xCOLD GASS results with outputs from hydrodynamic and semi-analytic models, we highlight the constraining power of cold gas scaling relations on models of galaxy formation. 
\end{abstract}


\section{Introduction}
\label{intro}

Much of galaxy evolution is regulated by the availability of gas, and the efficiency of the star formation process out of this material. For example, the shape, tightness and redshift evolution of the main sequence of star-forming galaxies in the {\edit star formation rate-stellar mass (SFR-\mstar)} plane can be explained by the availability of cold gas through inflows, the efficiency of the star formation process, and the balancing power of feedback \citep[e.g.][]{bouche10,lilly13,tacconi13,tacconi17,sargent14,saintonge16}. The cold atomic and molecular gas in the interstellar medium of galaxies is not only intimately linked to star formation, it is also an excellent probe of the larger environment and of evolutionary mechanisms. 

While initially mostly restricted to particularly luminous or nearby galaxies \citep[e.g.][]{sanders85,radford91,young95,solomon97}, over the past two decades samples of galaxies with integrated molecular line observations have grown to include in the local universe normal, non-interacting spiral galaxies \citep{braine93,sage93}, cluster spiral galaxies \citep{kenney88,boselli97}, early-type galaxies \citep{combes07,krips10,young11}, galaxies with active nuclei \citep{helfer93,sakamoto99,garcia03}, and isolated galaxies \citep{lisenfeld11}. Improvement in instrument sensitivities and bandwidth have also made possible the investigation of the molecular gas contents of galaxies far beyond the local Universe \citep{tacconi10,daddi10,geach11,magdis12b,magnellisaintonge,bauermeister13}. 

{\edit The COLD GASS survey \citep[CO Legacy Database for GASS, ][]{COLDGASS1}} was designed to provide a cohesive picture of molecular gas across the local galaxy population with \mstar$>10^{10}$\msun. Unlike all the studies mentioned above, the sample was selected purely by redshift and stellar mass rather than targeting specific classes of galaxies and, with 366 galaxies observed as part of a cohesive observing campaign, is homogenous and large enough to statistically characterise scaling relations and their scatter. Significant results from COLD GASS include the demonstration that star formation efficiency varies systematically across the galaxy population \citep{COLDGASS2,saintonge12}, and that the position of galaxies in the SFR-\mstar\ plane is driven by their gas contents and the varying star formation efficiency \citep{saintonge16}.  Because the COLD GASS sample is large and unbiased, it serves as the perfect reference for studies of particular galaxy populations (e.g. AGN hosts, interacting galaxies, early-type galaxies) and has been extensively used as such \citep[e.g.][]{fumagalli12,kirkpatrick14,bothwell14,stanway15,shimizu15,amorin16,alatalo16,yesuf17}. It is also an ideal $z=0$ reference point for studies of molecular gas at higher redshifts \citep[e.g.][]{combes13,troncoso14,seko16,genzel15,tacconi17}, and provides powerful constraints for theoretical models and numerical simulations \citep[e.g.][]{lagos11,genel14,popping15,lagos15,dave17}. 

Until now, COLD GASS could only provide information for the relatively massive galaxy population (\mstar$>10^{10}$\msun). Recognising the need to understand the link between gas, star formation, and global galaxy properties in lower mass galaxies, we launched a second IRAM-30m large programme to extend the sample down to stellar masses of $10^9$\msun. We present here the combination of the two surveys, now collectively referred to as xCOLD GASS for ``extended COLD GASS"; it contains IRAM-30m CO(1-0) measurements as well as a wide range of measured global properties for 532 galaxies spanning the entire SFR-\mstar\ plane at \mstar$>10^9$\msun. 

This paper presents the full catalog of IRAM-30m measurements, including both CO(1-0) and CO(2-1) line fluxes, complementary APEX CO(2-1) observations, and a catalog of global galaxy measurements derived from GALEX, SDSS, and WISE data products. We then use the xCOLD GASS sample to derive a robust CO luminosity function, determine the CO(2-1)/CO(1-0) excitation correction, present the extended scaling relations between gas fraction, depletion timescales and global galaxy properties, and conclude by discussing the key role of molecular gas observations in understanding the multi-scale nature of the star formation process, and in providing fresh and important constraints for models of galaxy formation. 
 
All rest-frame and derived quantities in this work assume a \citet{chabrier03} initial mass function, and a cosmology with $H_0=70$\kms\ Mpc$^{-1}$, $\Omega_m=0.3$ and $\Omega_{\Lambda}=0.7$.

\section{Sample and observations}
\label{data}

\subsection{Two IRAM large programmes}

The xCOLD GASS sample was assembled over the course of two large programs at the IRAM 30m telescope. The initial COLD GASS survey targeted 366 galaxies with $M_{\ast}>10^{10} M_{\odot}$ and $0.025<z<0.050$. The sample was selected randomly out of the complete parent sample of SDSS galaxies within the ALFALFA footprint matching these criteria, making it unbiased and representative of the local galaxy population.  A thorough description of the sample selection, survey strategy and scientific motivation is given in \citet{COLDGASS1}. 

The original COLD GASS survey was followed by a second effort, to extend the sample in the stellar mass range of $10^{9}<M_{\ast}/M_{\odot}<10^{10}$; we refer to this second survey as COLD GASS-low. Since the predicted CO luminosities of these galaxies are lower, the redshift range was lowered to $0.01<z<0.02$ for ease of detection. In this redshift range, the SDSS spectroscopic sample is complete for galaxies with $M_{\ast}>10^{9} M_{\odot}$, and the angular sizes of these lower mass galaxies are small enough that most of the CO flux can be recovered with a single pointing of the IRAM 30m telescope and a small aperture correction. A random subsample of {\edit 166 galaxies} from the SDSS parent sample of 764 galaxies was targeted with IRAM.

\begin{figure}
\epsscale{1.2}
\plotone{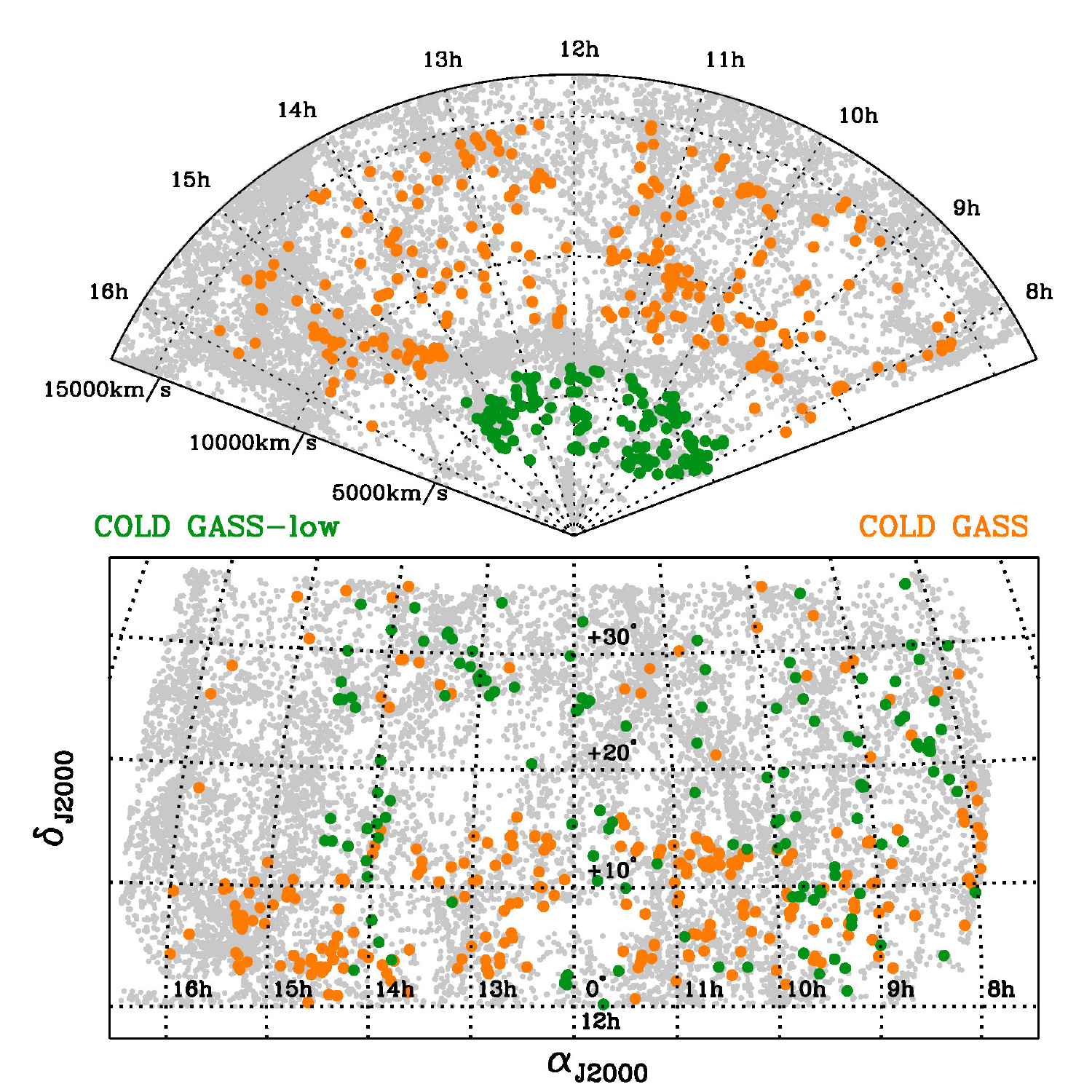}
\caption{Distribution of the ``Spring" component of the sample ($8{\rm h}<\alpha_{J2000}<16.0{\rm h}$) in the redshift-right ascension plane (top) and as projected on the sky (bottom).  \label{springsky}}
\end{figure}

\begin{figure}
\epsscale{1.2}
\plotone{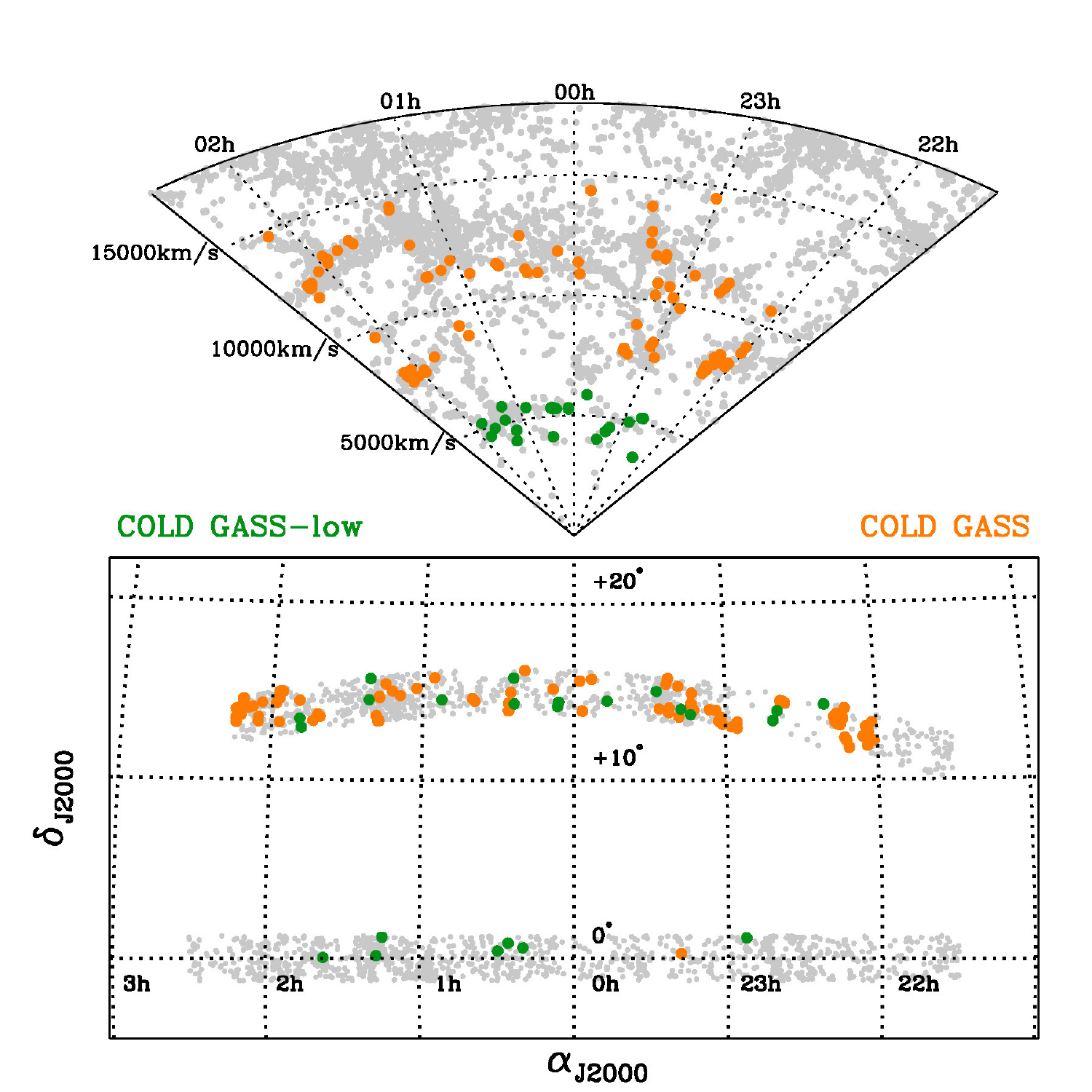}
\caption{Distribution of the ``Fall" component of the sample ($22{\rm h}<\alpha_{J2000}<2.5{\rm h}$) in the redshift-right ascension plane (top) and as projected on the sky (bottom).  \label{fallsky}}
\end{figure}

The combined xCOLD GASS sample therefore contains 532 galaxies with IRAM-30m CO(1-0) observations. The distribution of these objects on the sky and in redshift space are shown in Figures \ref{springsky} and \ref{fallsky}.  The redshift range of each survey is such that all sources could be observed with a single frequency tuning. Furthermore, the large footprint of the survey meant that observations could be performed at almost any time (and under almost any weather conditions), making the programmes ideal for pool observing, while the high density of sources on the sky implied that nearby sources could be observed without the need for major re-pointing/focussing of the telescope. These three elements were critical in enabling a very high observing efficiency and the assembly of a large sample (532 galaxies over a total of $\sim$950 hours of observing time). 

Both the sample selection and the observing strategy make xCOLD GASS the ideal sample to build scaling relations and serve as the benchmark for galaxy evolution studies. Such key features are:  (1) the representative nature of the sample, being purely mass-selected with no additional cuts on quantities such star formation rate, morphology or infrared luminosity, (2) the size of the sample which allows to define both mean scaling relations and any scatter or third-parameter dependencies,  (3) the homogeneity of the CO measurements and the strict upper limits set in the case of non-detections, and (4) the large dynamic range of the various physical properties under consideration (e.g. 2.5 dex in stellar mass, and metallicities ranging from 0.2 to 1.1$Z_{\odot}$).

\subsection{Galaxy properties and ancillary observations}
\label{galprops}

\begin{figure*}
\epsscale{1.2}
\plotone{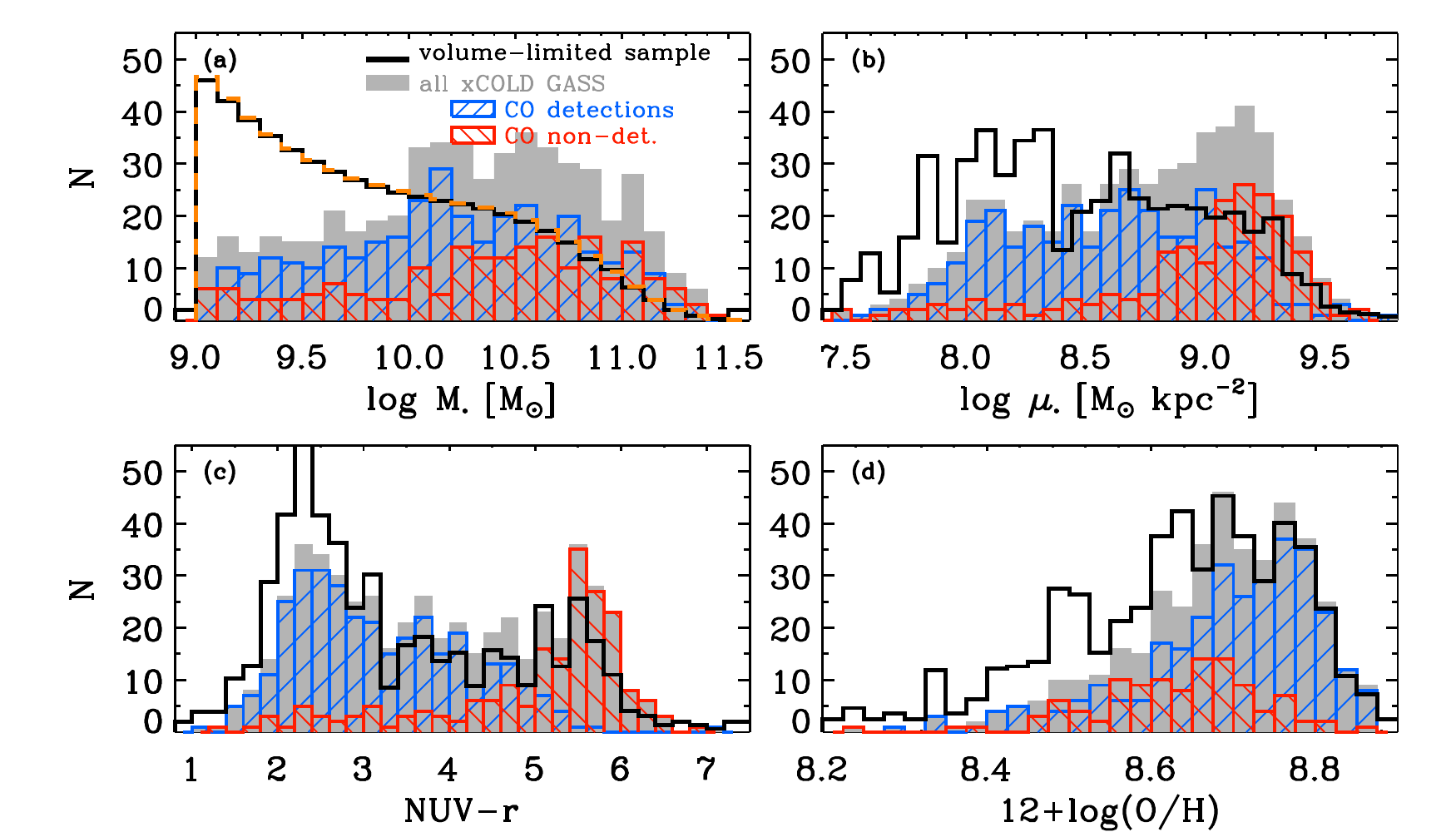}
\caption{Distributions of some key parameters across the xCOLD GASS sample: (a) stellar mass, (b) stellar mass surface density, (c) \nuvr\ color and (d) gas-phase metallicity calculated using the O3N2 calibration of \citet{pettini04}. In all panels, the filled gray histogram shows the distribution for the 532 xCOLD GASS objects. Galaxies with CO(1-0) detection and non-detection are shown separately as the blue and red histograms, respectively. The solid black line shows the distribution of the xCOLD GASS sample after weights are applied to correct for the flat $\log M_{\ast}$ distribution in the observed samples; this matches well the stellar mass function as shown as the orange dashed line in panel (a). Using the weighting, the xCOLD GASS sample can therefore be considered to be volume-limited. \label{histo}}
\end{figure*}

For each survey, the sample was randomly selected out of the SDSS parent sample to have a flat $\log M_{\odot}$ distribution to ensure an even sampling of the stellar mass parameter space. However, since the underlying stellar mass distribution of the full sample from SDSS is very well characterised, we can easily correct for this ``mass bias" \citep{GASS1}.  As a starting point, we construct the expected mass distribution of a purely volume-limited sample of 532 galaxies based on the \citet{baldry12} stellar mass function; this is the orange dashed line in Fig. \ref{histo}a, to be compared with the actual mass distribution of the xCOLD GASS sample, shown as the filled gray histogram. We assign as a statistical weight to each galaxy within $\log M_{\ast}$ bins of 0.1 dex in width the ratio between the number of galaxies expected from the stellar mass function and the number of objects in the xCOLD GASS sample (i.e. the weight as a function of $\log M_{\ast}$ is the ratio between the orange dashed histogram and the filled gray histogram of Fig. \ref{histo}a). To illustrate the impact of this weighting on other key parameters, Fig. \ref{histo} shows the difference between the observed distribution of stellar mass surface density, NUV$-r$ color and metallicity for the xCOLD GASS sample before and after the weights are taken into consideration (filled gray and black solid line histograms, respectively). Any scaling relation or mean quantity presented in this paper takes these weights into account, as they make the xCOLD GASS sample volume-limited. 

To derive accurate star formation rates for all xCOLD GASS galaxies, photometry is extracted from the WISE and GALEX survey databases. The IR and UV components of the total SFR are calculated and combined following exactly the method described in \citet{janowiecki17}. We also extract information from the SDSS DR7 database to provide us with information about the structural properties and chemical composition of the galaxies.  For all the xCOLD GASS objects, the following key parameters are given in Table \ref{sampleparams}: 
\begin{enumerate}
\item \texttt{GASS ID:} Catalog ID in the GASS survey. Galaxies with six digit IDs are part of COLD GASS-low
\item $\mathtt{\alpha_{J2000}}$: Right ascension of the SDSS object, in decimal degrees
\item $\mathtt{\delta_{J2000}}$: Declination of the SDSS object, in decimal degrees
\item \texttt{redshift:} SDSS spectroscopic redshift
\item \texttt{stellar mass:} Stellar mass from the SDSS DR7 MPA/JHU catalog\footnote{\url{http://home.strw.leidenuniv.nl/~jarle/SDSS/}}
\item \texttt{stellar mass surface density:} Calculated as $\mu_{\ast}=M_{\ast} (2 \pi r_{50,z}^2)^{-1}$, where $r_{50,z}$ is the radius encompassing 50\% of the $z$-band flux, in kpc. Galaxies with $\log \mu_{\ast}>8.7$ are considered to have a significant stellar bulge. 
\item \texttt{concentration index:} Ratio of the $r$-band Petrosian radii encompassing 90\% and 50\% of the flux,  $C=r_{90,r}/r_{50,r}$. The concentration index can be regarded as a proxy for the bulge-to-disc ratio \citep{weinmann09}, with $C>2.5$ being bulge-dominated galaxies. 
\item \texttt{effective radius:} SDSS $r$-band major axis effective radius, in kpc. 
\item \texttt{NUV-r color:} Calculated from {\edit resolution-matched} SDSS and GALEX photometry \citep[see ][]{wang10}, and corrected for Galactic extinction using the prescription of \citet{wyder07}. 
\item \texttt{star formation rate:} Calculated using an ``SFR ladder" technique as described in \citet{janowiecki17}. 
\item \texttt{metallicity:} Gas-phase metallicity 12$+\log$O/H using both the {\sc[NII]}/H$_{\alpha}$ and {\sc[OIII]}/H$_{\beta}$ line ratios and the calibration of \citet{pettini04}, for the galaxies classified as star-forming or composite in the BPT diagram (see next column). For galaxies with either weak emission lines or AGN activity, the metallicity is inferred from the mass-metallicity relation derived by \citet{kewley08} to be consistent with the strong line calibration of \citet{pettini04}. 
\item \texttt{BPT classification:} Spectral classification based on the BPT diagram,  0: inactive, 1: star forming, 2: composite, 3: AGN/LINER, 4: Seyfert, -1: undetermined. 
\end{enumerate}

\begin{deluxetable*}{lccccccccccc}
\tabletypesize{\scriptsize}
\tablecaption{Identifiers and properties of the xCOLD GASS galaxies \label{sampleparams}}
\tablehead{
\colhead{ID} & \colhead{$\alpha_{J2000}$} & \colhead{$\delta_{J2000}$} & \colhead{$z_{spec}$} & \colhead{$\log M_{\ast}$} & \colhead{$\log \mu_{\ast}$} &  \colhead{C} & \colhead{$r_{50}$} &  \colhead{NUV-$r$} & 
\colhead{SFR} & \colhead{12$+\log$O/H} & \colhead{BPT class} \\ 
 &  \colhead{[deg]}  & \colhead{[deg]} &  &  \colhead{$[\log M_{\odot}]$} &  \colhead{$[\log M_{\odot} kpc^{-2}]$} &  &  \colhead{[kpc]} & & 
 \colhead{$[M_{\odot} yr^{-1}]$} &  &  }
\startdata
124028 &   1.50998 &  14.41746 & 0.0176 &  9.11 &   7.71  &  2.39  &  2.47  &  2.96 &   0.14  &    8.48 &  1 \\
124012 &   1.62208 &  14.18237 & 0.0178 &  9.74 &   8.27  &  2.42  &  2.42  &  4.37 &   0.16  &    8.62 &  0 \\
 11956 &   2.08654 &  15.15601 & 0.0395 & 10.09 &   8.55  &  2.15  &  2.42  &  3.04 &   0.47  &    8.72 &  1 \\
 12025 &   4.89396 &  16.20418 & 0.0366 & 10.84 &   9.19  &  3.03  &  2.85  &  5.93 &   0.21  &    8.77 &  3 \\
124006 &   4.94721 &   0.59079 & 0.0177 &  9.75 &   8.76  &  3.15  &  1.21  &  3.57 &   0.55  &    8.62 &  4 \\
124000 &   5.93277 &  14.30672 & 0.0179 &  9.41 &   7.96  &  2.22  &  2.60  &  2.62 &   0.22  &    8.58 &  1 \\
124010 &   5.99747 &  15.77045 & 0.0177 &  9.93 &   8.18  &  2.14  &  4.12  &  2.04 &   2.94  &    8.80 &  1 \\
 12002 &   6.26670 &  14.97091 & 0.0367 & 10.48 &   9.47  &  3.17  &  1.32  &  6.25 &   0.01  &    8.76 & -1 \\
124004 &   6.39337 &   0.84685 & 0.0178 &  9.31 &   1.39  &  1.00  & \nodata  &  2.18 &   0.33  &    8.72 &  1 \\
 11989 &   6.49538 &  13.92940 & 0.0419 & 10.69 &   9.25  &  3.02  &  2.20  &  5.79 &   0.04  &    8.77 &  3 \\
 11994 &   6.49573 &  14.34799 & 0.0373 & 10.17 &   8.12  &  2.43  &  4.82  &  2.53 &   0.91  &    8.70 &  1 \\
124003 &   7.40992 &   0.41037 & 0.0138 &  9.22 &   7.76  &  1.95  &  2.43  &  1.82 &   0.30  &    8.48 &  1 \\
 27167 &   9.84028 &  14.46986 & 0.0380 & 10.37 &   9.21  &  2.77  &  1.62  &  4.48 &   0.16  &    8.75 &  3 \\
  3189 &  10.09787 &  14.61375 & 0.0384 & 10.05 &   7.99  &  1.96  &  4.54  &  2.77 &   2.37  &    8.69 &  0 \\
124027 &  13.11632 &  14.51829 & 0.0182 &  9.78 &   8.03  &  2.11  &  4.09  &  2.91 &   0.53  &    8.70 &  1 \\
  3261 &  13.88588 &  15.77582 & 0.0375 & 10.08 &   8.63  &  2.54  &  2.25  &  2.63 &   1.49  &    8.73 &  1 \\
  3318 &  15.65954 &  15.16852 & 0.0397 & 10.53 &   9.04  &  3.05  &  2.29  &  5.73 &   0.07  &    8.76 &  0 \\
  3439 &  17.27486 &  14.75578 & 0.0386 & 10.35 &   8.84  &  2.90  &  2.72  &  3.05 &   0.52  &    8.74 &  3 \\
  3465 &  18.09092 &  15.01085 & 0.0292 & 10.19 &   8.98  &  2.89  &  1.80  &  3.63 &   0.22  &    8.72 &  3 \\
101025 &  18.65576 &   1.18149 & 0.0154 &  9.84 &   8.02  &  1.71  &  3.96  &  2.84 &   0.45  &    8.76 &  1 \\
  3645 &  18.75730 &  15.41350 & 0.0307 & 10.33 &   8.98  &  2.71  &  1.98  &  3.97 &   0.15  &    8.74 &  3 \\
101021 &  19.22327 &   0.15313 & 0.0190 &  9.22 &   8.00  &  2.19  &  1.93  &  2.33 &   0.28  &    8.67 &  1 \\
  3509 &  19.29855 &  13.34094 & 0.0484 & 10.81 &   9.26  &  3.11  &  2.54  &  4.14 &   1.24  &    8.75 &  2 \\
  3524 &  19.31707 &  14.62238 & 0.0380 & 10.73 &   9.42  &  2.71  &  2.01  &  5.15 &   0.32  &    8.68 &  2 \\
  3519 &  19.36713 &  14.70443 & 0.0427 & 10.74 &   8.71  &  2.20  &  4.48  &  3.68 &   3.01  &    8.79 &  1 \\
  3505 &  19.44485 &  13.32348 & 0.0479 & 10.21 &   8.91  &  3.30  &  1.87  &  4.92 &   0.25  &    8.72 & -1 \\
  3504 &  19.59767 &  13.62457 & 0.0380 & 10.16 &   7.98  &  1.84  &  6.43  &  2.85 &   0.74  &    8.71 &  0 \\
101007 &  20.28890 &  15.69477 & 0.0171 &  9.45 &   7.88  &  1.75  &  3.01  &  2.26 &   0.54  &    8.64 &  1 \\
101004 &  20.37777 &  14.50498 & 0.0140 &  9.92 &   8.78  &  2.30  &  1.67  &  2.84 &   0.63  &    8.79 &  1 \\
101037 &  24.40472 &   0.04017 & 0.0164 &  9.31 &   8.00  &  2.48  &  2.14  &  2.14 &   0.32  &    8.60 &  1 
\enddata
\tablecomments{The full version of this table, including all 532 xCOLD GASS galaxies, is available online. A detailed description of this table's contents is given in Section \ref{galprops}.}
\end{deluxetable*}

\subsection{xGASS: The atomic gas survey}

Just like in the case of the original COLD GASS, the new COLD GASS-low survey was accompanied by a sister program at the Arecibo telescope (PI: B. Catinella). When available, HI masses are taken from the ALFALFA blind survey catalog \citep{alfalfa1,haynes11}, else they are observed with Arecibo as part of this programme. 

The GASS survey measured the \hi\ content for \about 700 galaxies with 
stellar masses greater than $10^{10}$ \msun\ and redshifts between
0.025 and 0.050 using the Arecibo radio telescope \citep{GASS1,GASS6,GASS8}. Galaxies were selected from SDSS DR6, with the additional
requirement on the sky footprint to be covered by projected GALEX MIS and ALFALFA surveys. Observations were limited to a gas fraction 
\mhi/\mstar\ between 2 and 5\%, depending on stellar mass. 

The low mass extension, GASS-low, targeted \about 200 galaxies with the same
stellar mass and redshift cuts as COLD GASS-low, selected from SDSS DR7
and lying within the ALFALFA 70\% footprint (Catinella et al. 2017). 
Observations were limited to gas fractions between 2 and 10\%.

In order to increase survey efficiency, galaxies with good detections 
from the ALFALFA survey were not re-observed, hence the observed sample
lacked \hi-rich objects. This was corrected by adding the correct proportion 
of randomly selected ALFALFA galaxies, based on the ALFALFA detection fraction
in each stellar mass bin. This procedure applied to both GASS and GASS-low
volumes yields the {\it xGASS representative sample}, hereafter simply
referred to as xGASS, which includes 1179 galaxies -- of these, 68\% are 
\hi\ detections. The GASS representative sample was revisited to take
advantage of more accurate ALFALFA detection fractions, now available
for larger volumes, and to maximise overlap with xCOLD GASS (see Catinella
et al 2017 for details). The overlap between xGASS and xCOLD GASS includes 477 galaxies; the \hi\
detection fraction for this subset is 73\%.

\subsection{IRAM 30m observations and data reduction}
\label{IRAMdata}

\begin{deluxetable*}{lcc}
\tabletypesize{\small}
\tablewidth{0pt}
\tablecaption{xCOLD GASS IRAM 30m observing parameters \label{IRAMparams}}
\tablehead{
 & \colhead{COLD GASS} & \colhead{COLD GASS-low}  }
\startdata
      Observing period & 2009 June - 2011 December & 2012 July - 2017 May   \\ 
      Frontend & EMIR & EMIR   \\ 
      Backend & WILMA \& 4MHz Filterbank & FTS  \\ 
      E090 frequency & 109.3-113.0 GHz & 107.4-115.2 GHz  \\ 
      E230 frequency & 220.7-224.4 GHz & 224.7-232.5 GHz \\
      Spectral resolution & 2 MHz & 0.1953 MHz \\
      Wobbler frequency / throw & 1 Hz / 180\arcsec & 1 Hz / 120\arcsec 
\enddata
\end{deluxetable*}

All xCOLD GASS observations of the CO(1-0) line were carried out at the IRAM 30m telescope using the Eight Mixer Receiver \citep[EMIR;][]{carter12}. In the 3mm band (E090), we can make use of two sidebands each with a bandwidth of 8GHz per linear polarisation. For each survey, the E090 band was tuned to a specific frequency which allowed us to detect the redshifted CO(1-0) line for all galaxies within the available bandwidth. The second band was tuned to a fixed frequency in the 1mm (E230) band to cover the redshifted CO(2-1) line which fell within the available 4 GHz bandwidth for 68\% of our sample.  All details regarding the set up of the instruments during both surveys are presented in Table \ref{IRAMparams}. 

All observations were done in wobbler-switching mode. For the initial survey, we used the Wideband Line Multiple Autocorrelator (WILMA) as the primary backend, with the 4MHz Filterbank as a backup. This allowed for simultaneous coverage of 4 GHz of bandwidth in each linear polarisation and for each band. For COLD GASS-low we were able to take advantage of the new Fast Fourier Transform Spectrometer (FTS), making it possible to record the full 8GHz of bandwidth from EMIR. The spectral resolution of the FTS is also a factor of ten higher.  The beam size of the telescope is 22\arcsec\ at 3mm and 11\arcsec\ at 1mm. 

The observing for xCOLD GASS took place between 2009 and 2015, with widely different atmospheric conditions. The strategy was to observe the galaxies predicted to be the most CO-luminous under poorer weather conditions, as these require a typical rms sensitivity of 1.5-2.0 mK per 20 \kms\ channel to achieve a detection of CO(1-0) with S/N$>5$.  When the precipitable water level was particularly low, we favoured the redder galaxies (predicted to be CO-faint) in order to achieve rms sensitivities of 0.8-1.0 mK in a reasonable amount of time (average of $\sim 2$ hour per target). The overall strategy was to observe a galaxy until either the CO(1-0) line was detected with S/N$>5$, or the rms noise was low enough to allow us to put a stringent upper limit on the gas fraction of $M_{H2}/M_{\ast}=1.5\%$ for COLD GASS and 2.5\% for COLD GASS-low. 

The IRAM data were all reduced with the CLASS software. Individual scans were baseline-subtracted (first order fit, {\edit excluding the spectral region within $\pm300$\kms\ of the expected line center based on the optical redshift}), visually examined to reject those with distorted baselines or anomalous features, and combined into a final averaged spectrum binned to a resolution of 20 \kms.  All the IRAM-30m spectra are shown in the Appendix, alongside the optical SDSS image of each galaxy.  A spectral window is defined for each emission line to match the observed line width. In cases where the CO line is undetected, the window is either set to the width of the HI line, or to a width of 300 \kms\ (200\kms\ for COLD GASS-low) if no HI information is available.The integrated line flux, $S_{CO,obs}$, is measured by adding the signal within this spectral window, and the standard deviation of the noise per 20 \kms\ channel, $\sigma_{rms}$ is measured outside of it.  Properties of the spectral lines, such as central velocity and width, are measured using a custom-made IDL interactive program following the technique described in e.g. \citet{springob07}, \citet{catinella07} and \citet{COLDGASS1}. For the galaxies with the highest $S/N$, a second width-fitting scheme is applied, allowing for careful error estimation as is required for Tully-Fisher work \citep{tiley16}.

Given the angular size of the galaxies, most of their flux can be recovered by a single pointing of the IRAM-30m telescope. However, to account for the larger angular size of some of the galaxies, we apply an aperture correction to all of the measured CO(1-0) line fluxes. The method presented in \citet{saintonge12} is used. In short, for each xCOLD GASS object, a model galaxy is created assuming an exponential molecular gas disk with a scale length equivalent to the radius enclosing half of the star formation rate. The model is given the inclination of the real galaxy, and then convolved with a Gaussian matching the properties of the IRAM beam. The aperture correction is the ratio between the flux of the model before and after this convolution. The median aperture correction across the xCOLD GASS sample is 1.17. {\edit We performed tests to ensure that the scaling relations presented later in this paper are not caused by inadequate aperture corrections by confirming that key quantities that should not depend on distance within our sample (such as molecular gas fraction and depletion timescale) are indeed uncorrelated.}

\subsection{IRAM catalog description}
\label{iramcat}
Based on the line fluxes, widths, and spectral noise measurements described above, we introduce the full catalog of CO(1-0) and CO(2-1) measurements for all 532 xCOLD GASS. As presented in Table \ref{COtab}, the catalog includes the following quantities: 
\begin{enumerate}
\item \texttt{GASS ID:} Catalog ID in the GASS survey. Galaxies with six digit IDs are part of COLD GASS-low. 
\item $\mathtt{\sigma_{CO10}}$: rms noise achieved around the CO(1-0) line, in spectral channels with width $\Delta w_{ch}=20$\kms.
\item \texttt{FlagCO10:} Detection flag for the CO(1-0) line, 1: detection, 2: non-detection. In the cases where the line is not detected, the line luminosities and molecular gas mass given in columns 7, 8 and 20 are $3\sigma$ upper limits. 
\item $\mathtt{S/N_{CO10}}$: Signal-to-noise ratio achieved in the CO(1-0) line, calculated as $S_{CO10,obs} / \epsilon_{obs}$, where 
\begin{equation}
\epsilon_{obs}=\frac{\sigma_{CO10} W50_{CO10}}{\sqrt{W50_{CO10} \Delta w_{ch}^{-1}}}, 
\label{eq:sigma}
\end{equation}
{\edit where $\Delta w_{ch}=20$\kms\ is the spectral channel width, $\sigma_{CO10}$ is the rms noise (see Col. 2) and $W50_{CO10}$ is the line width (see Col. 10).}
\item $\mathtt{S_{CO10,obs}}$: Integrated CO(1-0) line flux within the IRAM beam, measured as described in Sec. \ref{IRAMdata}. The values are given in units of Jy~\kms, after applying the point source sensitivity factors recommended for EMIR\footnote{\url{http://www.iram.es/IRAMES/mainWiki/EmirforAstronomers}}
\item $\mathtt{S_{CO10,cor}}$: Total CO(1-0) line flux, calculated from $S_{CO10,obs}$ and the galaxy-specific aperture correction, as described in Sec. \ref{IRAMdata}. This is the flux value recommended to be used to infer the total molecular gas mass of the xCOLD GASS galaxies. 
\item $\mathtt{L^{\prime}_{CO10,obs}}$: Beam-integrated CO(1-0) line luminosity in units of K~\kms~pc$^{2}$ calculated following \citet{solomon97}: 
\begin{equation}
L^{\prime}_{CO10,obs}=3.25\times10^{7}S_{CO10,obs}\nu_{obs}^{-2}D_L^2(1+z)^{-3},
\end{equation}
where $\nu_{obs}$ is the observed frequency of the CO(1-0) line in GHz and $D_L$ is the luminosity distance in units of Mpc. The error includes the measurement uncertainty and the 8\% flux calibration error at these frequencies. Uncertainties on pointing and redshift are negligible in comparison and not included \citep{COLDGASS1}. If FlagCO10$=2$ (see Col. 3), then the value is a $3\sigma$ upper limit, where $\sigma$ is calculated using Eq. \ref{eq:sigma} assuming that $W50_{CO10}=W50_{HI}$ when an HI detection is available, otherwise constant values of 300 and 200 \kms\ for galaxies with \mstar\ greater and lower than $10^{10}$\msun, respectively. 
\item $\mathtt{L^{\prime}_{CO10,cor}}$: Total CO(1-0) line luminosity, calculated from $L^{\prime}_{CO10,obs}$ and the aperture correction.  The error includes the measurement uncertainty, the 8\% flux calibration error, and the 15\% uncertainty on the aperture correction \citep{saintonge12}. If FlagCO10$=2$ (see Col. 3), then the value is a $3\sigma$ upper limit.
\item $\mathtt{z_{CO10}}$: Redshift of the galaxy based on the central velocity of the CO(1-0) line
\item $\mathtt{W50_{CO10}}$: Full width at half maximum of the CO(1-0) line in \kms, calculated using the technique of \citet{springob05}, which is based on fitting linear slopes to the two sides of the emission line and finding the width at half maximum along these fits.
\item $\mathtt{W50_{TFR}}$: Full width at half maximum of the CO(1-0) line in \kms, calculated for the galaxies with the highest S/N for the purpose of Tully-Fisher studies by fitting the emission lines with a Gaussian Double Peak function \citep{tiley16}. 
\item $\mathtt{\sigma_{CO21}}$: rms noise achieved around the CO(2-1) line, in spectral channels with width $\Delta w_{ch}=20$\kms. Since the integration times were set by the requirements to detect the CO(1-0) line, the depth of these observations is variable as these frequency are far more susceptible to elevated levels of atmospheric water vapor. 
\item \texttt{FlagCO21:} Detection flag for the CO(2-1) line, 0: not targeted, 1: detection, 2: non-detection. In the cases where the line is not detected, the CO(2-1) line luminosity in Col. 16 is a $3\sigma$ upper limit. 
\item $\mathtt{S/N_{CO21}}$: Signal-to-noise ratio achieved in the CO(2-1) line, calculated as described for Column 3.
\item $\mathtt{S_{CO21,obs}}$: Integrated CO(2-1) line flux within the IRAM beam, in Jy~\kms. Note that the beamsize at these frequencies is half the size of that at the frequency of the CO(1-0) line. For this reason, the aperture correction to extrapolate the CO(2-1) to a total value are large and more uncertain, therefore we only provide observed beam quantities for this line. 
\item $\mathtt{L^{\prime}_{CO21,obs}}$: Beam-integrated CO(2-1) line luminosity, in K~\kms~pc$^{2}$ calculated as described in Col. 7.
\item $\mathtt{z_{CO21}}$: Redshift of the galaxy based on the central velocity of the CO(2-1) line
\item $\mathtt{W50_{CO21}}$: Full width at half maximum of the CO(2-1) line in \kms, calculated as in Col. 10. 
\item $\mathtt{\alpha_{CO}}$: Recommended value for the CO-to-H$_2$ conversion factor, calculated using the function calibrated by \citet{accurso17b}.  This is a metallicity-dependent conversion function, with a second order dependence on the offset of a galaxy from the star-forming main sequence. 
\item $\mathtt{\log M_{H2}}$: Total molecular gas mass, including the Helium contribution, calculated from the total CO(1-0) line luminosity presented in Col. 8, and the conversion function in Col. 19 as $M_{H2}=\alpha_{CO}(Z,{\rm SSFR}) L^{\prime}_{CO10,cor}$. The error provided includes the uncertainty on the CO(1-0) line luminosity (see Col. 8), and the 35\% uncertainty on the $\alpha_{CO}$ conversion function as determined by \citet{accurso17b}. If FlagCO10$=2$ (see Col. 3), indicating a non-detection of the CO(1-0) line, then the value given is a $3\sigma$ upper limit. 
\end{enumerate}

The full xCOLD GASS catalog, including all quantities presented in Tables \ref{sampleparams} and \ref{COtab}, as well as all the IRAM CO spectra and SDSS images presented in the Appendix can be retrieved from \url{http://www.star.ucl.ac.uk/xCOLDGASS}.

\subsection{APEX CO(2-1) observations}
\label{apexdata}

\begin{figure*}
\epsscale{1.1}
\plotone{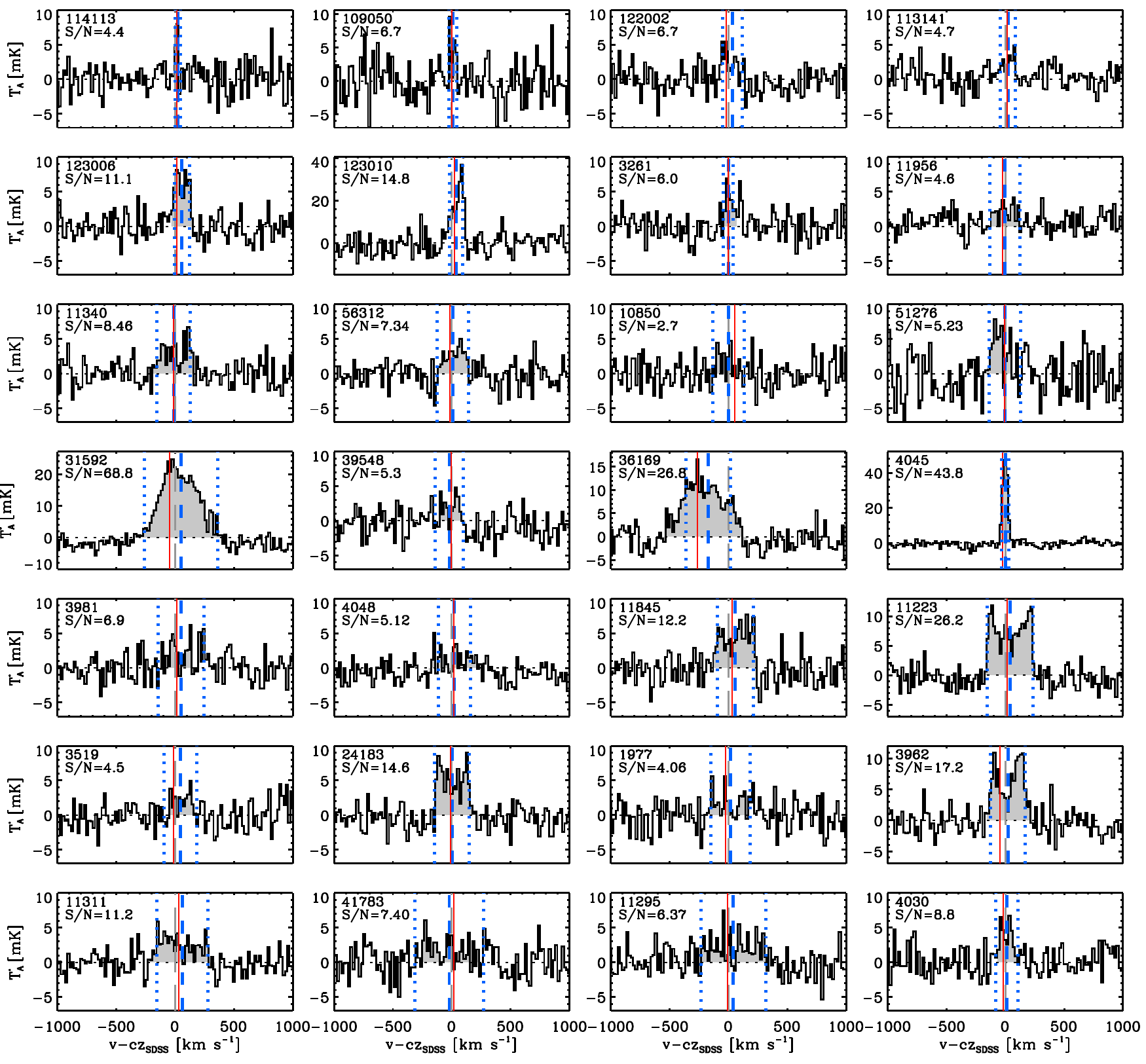}
\caption{APEX CO(2-1) spectra of xCOLD GASS objects. Shown are the 28 galaxies from Table \ref{APEXtable} with detections of the CO(2-1) line. In each panel, the red solid line is the expected line centre based on the SDSS optical redshift, while the dashed and dotted blue vertical lines are the measured CO line centre and width. {\edit The grey shaded areas represent the region of the spectra over which we integrated to calculate the total line flux.} \label{APEXspectra}}
\end{figure*}

The CO(1-0) emission line is only one of the many tracers available to measure the mass of molecular gas in galaxies. At higher redshifts, it is more common to observe higher $J$ transitions of this same molecule, as for example, lines such as CO(2-1) and CO(3-2) fall into the 3mm atmospheric window at $z\sim1$ and $z\sim3$, respectively. Having both CO(1-0) and CO(2-1) observations for 68\% of the xCOLD GASS sample, we would be in a good position to investigate how the luminosity ratio, $r_{21}\equiv L_{CO(2-1)}/L_{CO(1-0)}$, varies as a function of global galaxy properties. This is important as a value of $r_{21}$ has to be assumed to convert an observed CO(2-1) flux into \mh. However, the beam size difference implies that the IRAM CO(1-0) and (2-1) observations do not probe the same area of the galaxies; any galaxy-to-galaxy variations in the (2-1)/(1-0) ratio measured with IRAM could be due either to gas excitation variations, or differences in the radial distribution of the gas. 

To help disentangle these effects, we have obtained additional CO(2-1) data with the APEX telescope. With a beam size of 27\arcsec\ at this frequency, these observations are a good match for the CO(1-0) fluxes measured with the IRAM 22\arcsec\ beam.  Observations were performed through the allocation of a total of 78 hours via both the ESO (proposal 091.B-0593) and Max-Planck channels. The APEX-1 receiver was used, tuned to the redshifted frequency of the CO(2-1) of each xCOLD GASS object, with a simple on-off observing mode. Integration times were set based on average observing conditions, and rms noise requirements  predicted from the measured IRAM CO(1-0), a standard value of $r_{21}=0.7$ and a Galactic conversion factor. The detection rate ($S/N>4$) is 85\%. The APEX data were reduced in CLASS using a procedure identical to that described above for the IRAM data. The reduced spectra for the 28 xCOLD GASS galaxies with an APEX CO(2-1) detection are shown in Figure \ref{APEXspectra}, with measured line properties presented in Table \ref{APEXtable}. 

\begin{deluxetable}{lcccc}
\tabletypesize{\small}
\tablewidth{0pt}
\tablecaption{Summary of APEX CO(2-1) observations \label{APEXtable}}
\tablehead{ \colhead{ID} & \colhead{rms} & \colhead{$S_{CO2-1}$} & \colhead{$L^{\prime}_{CO2-1}$} & \colhead{S/N} \\ 
 & \colhead{[mK]} & \colhead{[Jy km/s]} & \colhead{[$10^8$ K km/s pc$^2$]} &   }
\startdata
 11956 & 1.42 &  18.41 &   3.29 &  4.69 \\
  3189 & 2.06                  & \nodata & \nodata & \nodata \\
  3261 & 1.70 &  13.38 &   2.15 &  6.00 \\
  3519 & 1.72 &  18.92 &   3.96 &  4.60 \\
  3962 & 1.79 &  92.82 &  19.45 & 17.27 \\
  4030 & 1.54 &  24.18 &   6.72 &  8.88 \\
  4048 & 1.28 &  22.11 &   4.34 &  5.13 \\
 24183 & 1.53 &  65.52 &  13.81 & 14.68 \\
 41783 & 1.75 &  48.36 &   7.59 &  7.41 \\
 39548 & 1.65 &  20.05 &   2.75 &  5.33 \\
 10850 & 1.49 &  10.22 &   1.47 &  2.71 \\
 11223 & 1.43 & 129.87 &  18.75 & 26.25 \\
 11311 & 1.31 &  53.04 &   7.05 & 11.29 \\
 11295 & 1.94 &  50.70 &   9.20 &  6.38 \\
 11270 & 1.65                  & \nodata & \nodata & \nodata \\
 11845 & 1.69 &  67.08 &  10.12 & 12.30 \\
  1977 & 1.83 &  20.94 &   2.12 &  4.07 \\
 11019 & 1.74                  & \nodata & \nodata & \nodata \\
 11340 & 1.70 &  41.34 &   6.05 &  8.46 \\
  4045 & 1.82 & 106.86 &   8.54 & 43.81 \\
  3981 & 1.39 &  23.13 &   4.59 &  6.95 \\
 51276 & 2.50 &  34.01 &   3.39 &  5.24 \\
 56312 & 1.64 &  30.89 &   4.16 &  7.35 \\
 41869 & 1.91                  & \nodata & \nodata & \nodata \\
 31592 & 1.50 & 368.16 &  89.77 & 68.90 \\
 36169 & 2.14 & 225.42 &  57.33 & 26.85 \\
109050 & 1.25 &  12.17 &   0.19 &  6.75 \\
114072 & 1.53                  & \nodata & \nodata & \nodata \\
122002 & 1.51 &  21.61 &   0.74 &  6.75 \\
111047 & 2.87                  & \nodata & \nodata & \nodata \\
123006 & 1.66 &  40.17 &   1.07 & 11.18 \\
123010 & 3.89 & 129.87 &   3.18 & 14.84 \\
114113 & 1.87 &   7.72 &   0.26 &  4.49 \\
113141 & 1.40 &  13.57 &   0.27 &  4.79 
\enddata
\end{deluxetable}

\section{Results}
\label{results}

\subsection{The CO luminosity function}

\begin{figure}
\epsscale{1.2}
\plotone{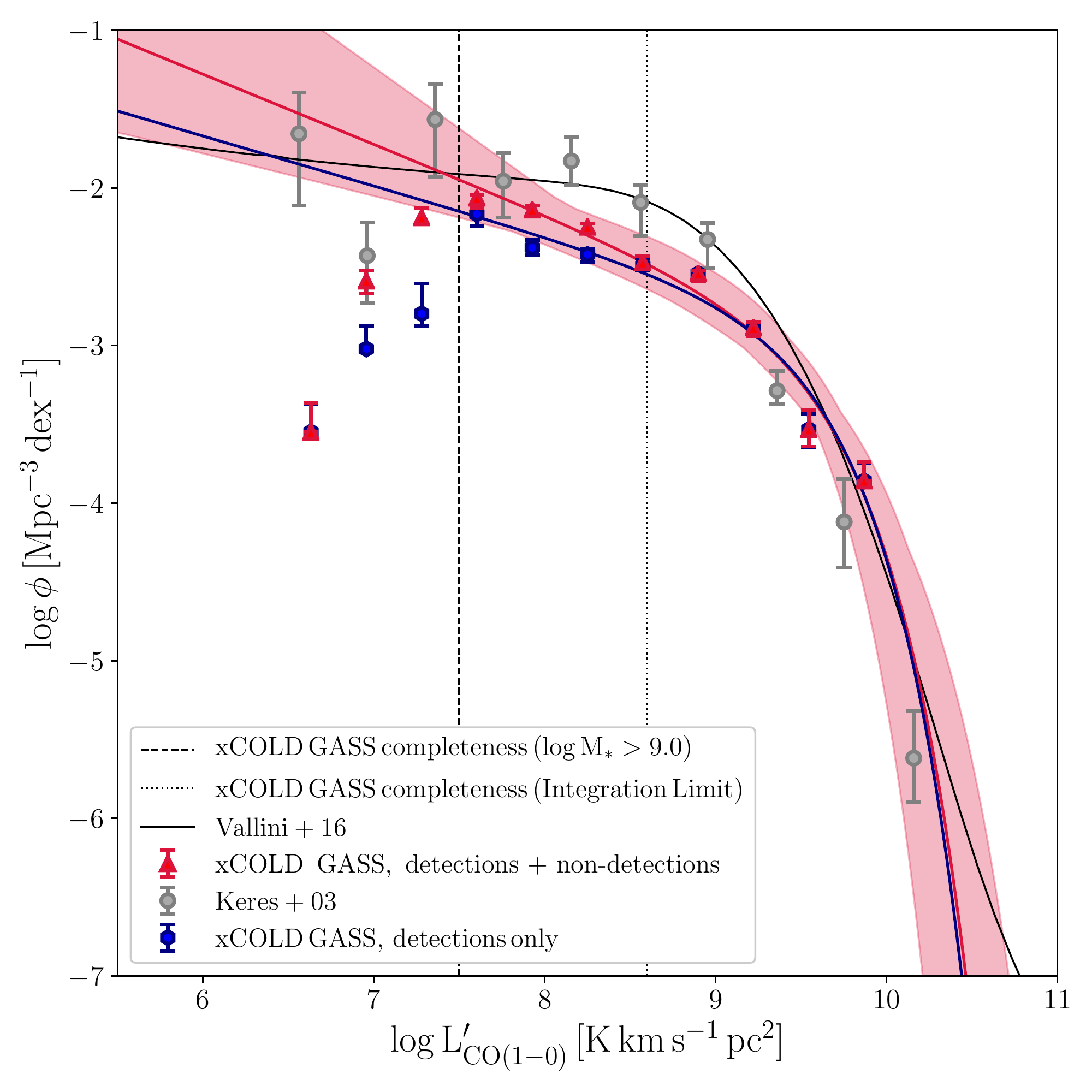}
\caption{The CO(1-0) luminosity from the xCOLD GASS sample where $\log L^{\prime}_{CO}$ is given in units of K~\kms~pc$^2$. The LF is shown if detections and the upper limits from non-detections are included (red triangles) and if only detections are used (blue hexagons). The best-fit Schechter function for the total xCOLD GASS sample where the data is complete is shown as a red line. For comparison the observed CO LF of local galaxies from \citet{keres03} (grey circles) and the empirical prediction for the CO LF using the $\log L^{\prime}_{CO}-L_{IR}$ conversion from \cite{vallini16} (solid black line) are also presented. The xCOLD GASS sample is complete down to $\log L^{\prime}_{CO} = 7.5$.}
  \label{fig:lum_func}
\end{figure}

\begin{deluxetable*}{lcccc}
\tabletypesize{\small}
\tablewidth{0pt}
\tablecaption{Best-fit Schechter function parameters for the xCOLD GASS CO luminosity functions \label{LFtable}}
\tablehead{ \colhead{ } & \colhead{$\phi^{\ast}$} & \colhead{$L^{\ast}_{\rm CO}$} & \colhead{$\alpha$}  \\ 
 & \colhead{[Mpc$^{-3}$]} & \colhead{[K \kms pc$^2$]} &    }
\startdata
Detections only & $(1.30\pm0.71) \times 10^{-3}$ & $(7.00 \pm 1.88)\times 10^9$ & $-1.13\pm0.05$ \\
Detections $+$ non-detections & $(9.84\pm5.41) \times 10^{-4}$ & $(7.77 \pm 2.11)\times 10^9$ & $-1.19\pm0.05$ 
\enddata
\end{deluxetable*}

The CO luminosity function (LF) at $z=0$ is a useful tool, which xCOLD GASS is particularly well suited to calibrate. Commonly used calibrations of the CO LF were derived either using samples biased towards more extreme starburst galaxies \citep{keres03,obreschkow09}, or indirectly using an empirical conversion between IR luminosities and CO line emission \citep{berta13,vallini16}. The xCOLD GASS sample, being only stellar mass-selected, provides a unique opportunity to obtain a more complete and direct view of the CO LF at $z=0$, as long as the completeness limits imposed by the stellar mass cutoff and the depth of the observations are taken into account.

While the sample was selected to have a flat stellar mass distribution (see Fig. \ref{histo}), because it is extracted from a volume-limited parent sample, we only have to apply the statistical weights described in Section  \ref{galprops} to recover a complete luminosity function (LF). We note that we do not attempt here to correct for the contribution of galaxies with \mstar$<10^9$\msun, and therefore produce a LF that accounts for the contributions of galaxies more massive that $10^9$\msun. However, given that there is a positive correlation between \mstar\ and $L^{\prime}_{CO}$, and given that the most CO-bright galaxies with \mstar$\sim10^9$\msun\ have $\log L^{\prime}_{CO}<7.5$, it is highly unlikely that lower mass galaxies contribute much if at all to the CO LF at $\log L^{\prime}_{CO}>7.5$.  There is a second completeness limit to consider, caused by the xCOLD GASS observing strategy. Because the sample is gas-fraction limited ({\edit i.e., the sensitivity limit is determined by the requirement to detect a fixed, constant gas mass fraction $M_{H_2}/$\mstar\ in all the galaxies}, see Sec. \ref{IRAMdata} and Fig. \ref{fH2data}), the depth in terms of CO luminosity varies by about an order of magnitude across the stellar mass range of the sample.  At the highest mass end of the sample, where the observations are shallowest in terms of CO luminosity, the gas fraction limit of the observations corresponds to $\log L^{\prime}_{CO}\sim 8.6$. 

The CO luminosity function for the xCOLD GASS sample is presented in Figure \ref{fig:lum_func}, with the estimated completeness limits due to stellar mass cut of the sample and the gas fraction integration limit shown by vertical lines at $\log L^{\prime}_{CO}\sim 7.5$ and $\log L^{\prime}_{CO}\sim 8.6$ respectively. The xCOLD GASS CO LF was calculated using the statistical weights described in Section \ref{galprops}, making the sample volume-limited. To address the completeness issue caused by the fixed gas fraction integration limit, we produce two LFs by treating the upper limits differently: in one case, we assign to the undetected galaxies their 3$\sigma$ upper limit on $\log L^{\prime}_{CO}$ (red symbols in Figure \ref{fig:lum_func}), and in the other we assign to them a value of $\log L^{\prime}_{CO}=0$ (blue symbols). As expected, these two LFs are identical above the completeness limit of $\log L^{\prime}_{CO}\sim 8.6$. The error on each point in the luminosity function was calculated using a bootstrap method.  In this procedure, a randomly selected sub-sample of 80\% of the full xCOLD GASS sample was selected, giving $N$ galaxies. For each of these $N$ galaxies, we assign a CO luminosity by randomly sampling a Gaussian centered around the measured CO luminosity of this galaxy and with a standard deviation corresponding to the 1$\sigma$ error. The CO LF for this sub-sample of $N$ galaxies is then constructed, and this procedure is repeated $N$ times. The final uncertainty on each data point in the LF determined from the full xCOLD GASS sample is the $1\sigma$ distribution at the given luminosity within the $N$ sub-sampled LFs produced by the bootstrap method.  These are the points and errors shown in Fig. \ref{fig:lum_func}, for both the methods of treating the non-detections. 

Figure \ref{fig:lum_func} illustrates that even in the $L^{\prime}_{CO}$ interval where the data is not affected by the stellar mass limit of the sample, the xCOLD GASS CO LF differs from the observed and empirically derived LFs from \cite{keres03} and \cite{vallini16}, respectively. The knee of the xCOLD GASS CO LF is at larger $L_{\mathrm{CO}}$, shifted downwards in number density and has a steeper slope. When the non-detections are assigned their 3$\sigma$ upper limits (red symbols),  the faint-end slope appears to be yet steeper. The true CO luminosity of many of these undetected galaxies is likely to be much lower than the $3\sigma$ upper limits used here, and we therefore expect the ``true" CO LF to lie somewhere between the functions determined with and without non-detections.

Finally, we fit the two xCOLD GASS CO luminosity functions with a Schechter function \citep{schechter76} over the luminosity interval where the stellar mass limit of the sample is not affecting the completeness ($\log L^{\prime}_{CO} > 7.5$). The bootstrap errors on the individual points of the LFs are considered, and the covariance matrix used to randomly sample the best-fit function. The best-fit Schechter functions to both LFs are shown in Fig. \ref{fig:lum_func}, with the range of possible fits within the uncertainties illustrated by the red shaded area for the case where both CO detections and non-detections are considers. The parameters of the best-fit Schechter function for both cases are given in Table \ref{LFtable}.

\subsection{Gas excitation}

\begin{figure}
\epsscale{1.1}
\plotone{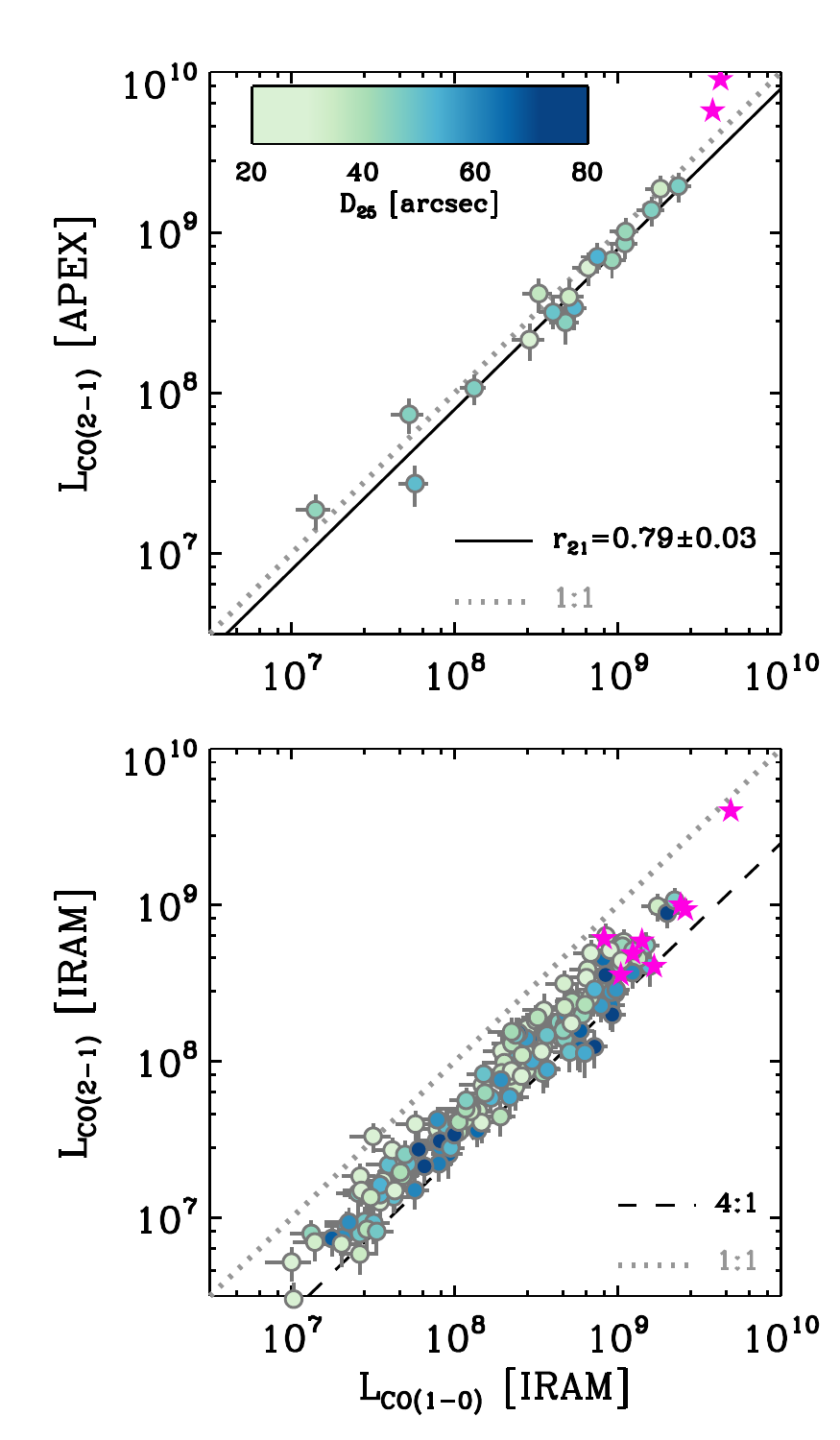}
\caption{Relation between observed CO(1-0) and CO(2-1) line luminosities, with the former having been observed through the 27\arcsec\ beam of the APEX telescope ({\it top}), or the 11\arcsec\ beam of the IRAM telescope ({\it bottom}).  In the top panel, the CO(1-0) luminosities have been corrected by a small aperture correction to account for the beam size difference, and the solid line shows the best fitting value for $r_{21}$. In the bottom panel, the dotted line is the 1:1 relation while the dashed line is a 4:1 relation. Galaxies are color-coded by their optical diameter, and the magenta stars represent merging systems. \label{r21plot}}
\end{figure}

In Figure \ref{r21plot} the line luminosities of the CO(1-0) and CO(2-1) lines are compared. When measurements are made within the same aperture, there are very little galaxy-to-galaxy variations in the integrated $r_{21}$ ratio. {\edit To achieve this comparison, we apply a small aperture correction to the IRAM CO(1-0) luminosities to account for the small difference in beam size with the APEX CO(2-1) measurements. These corrections, based on the technique described in Sec.~\ref{IRAMdata}, are in the range of 2-10\%.} 

Across the joint IRAM-APEX sample, the mean value of the luminosity ratio is $r_{21}=0.79\pm0.03$. The linear scatter around this value is 0.23, and reduces to 0.15 if considering only galaxies with $L_{CO(1-0)}>10^8$ K~\kms~pc$^2$. The xCOLD GASS value of $r_{21}\sim0.8$ corresponds to an excitation temperature of 10K, assuming that the gas is optically thick \citep{leroy13}. This value of $r_{21}$ is slightly larger than the values of 0.5-0.7 reported from resolved observations of the discs of nearby star-forming spiral galaxies \citep[e.g.][]{leroy13,rosolowsky15}. Higher values still of $r_{21}\sim1$ are usually reported for the nuclei of galaxies \citep[e.g.][]{braine92,sakamoto95,leroy09}. {\edit In Fig.~\ref{r21plot} we highlight the position of merging systems found within the xCOLD GASS dataset. We may expect the gas in these systems to have higher excitation and/or to be more centrally concentrated as seen in some other nearby systems with detailed multi-transition CO observations \citep[e.g.][]{saito17}, although the physical conditions of the gas are likely a function of the specific geometry and evolution state of the merger \citep{ueda12}. While there are not enough merging systems within the xCOLD GASS sample to study these subtleties in detail, it makes an excellent comparison sample for observing programs specifically targeting such dynamically active systems.} 

\subsection{Gas fraction scaling relations}

\begin{figure*}
\epsscale{1.05}
\plottwo{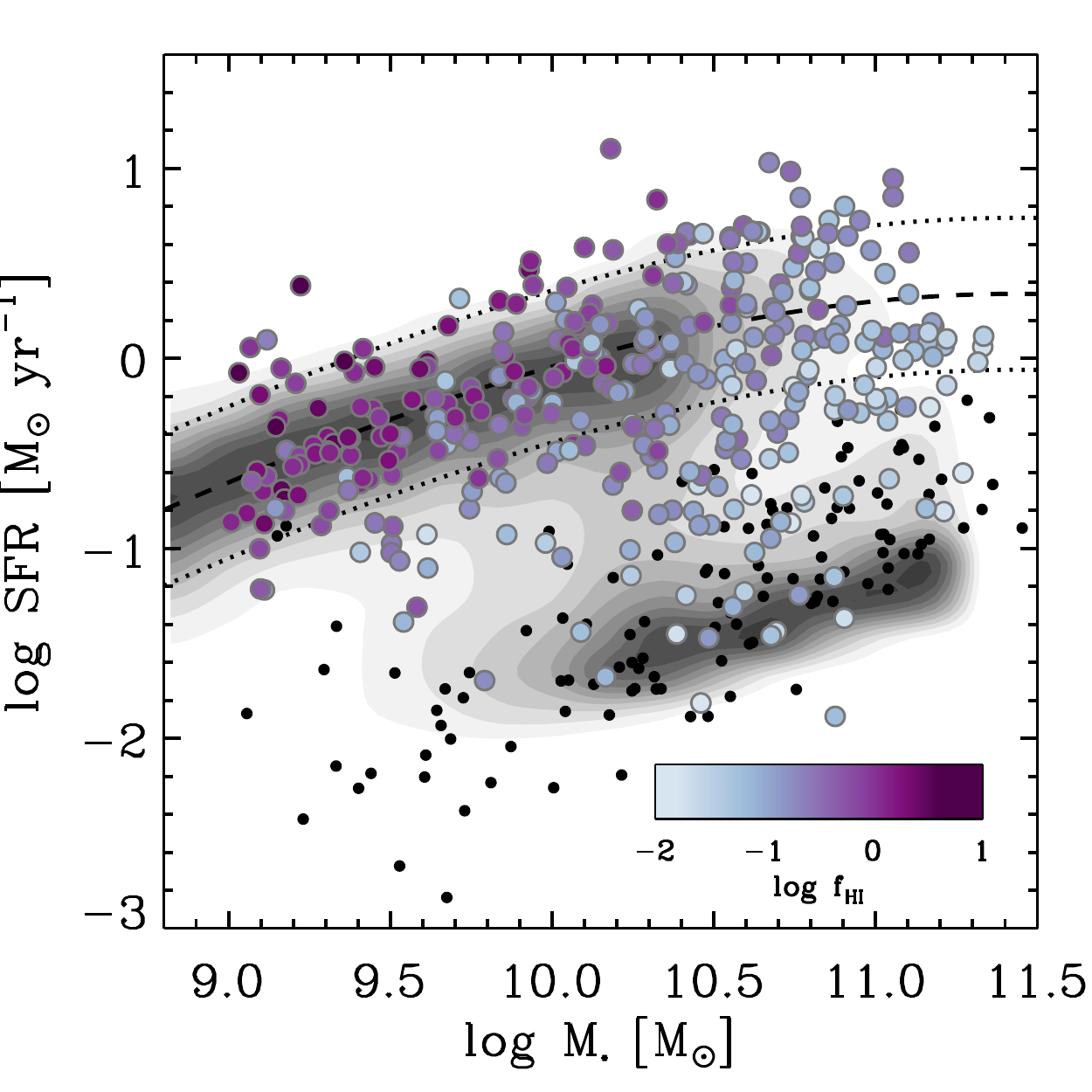}{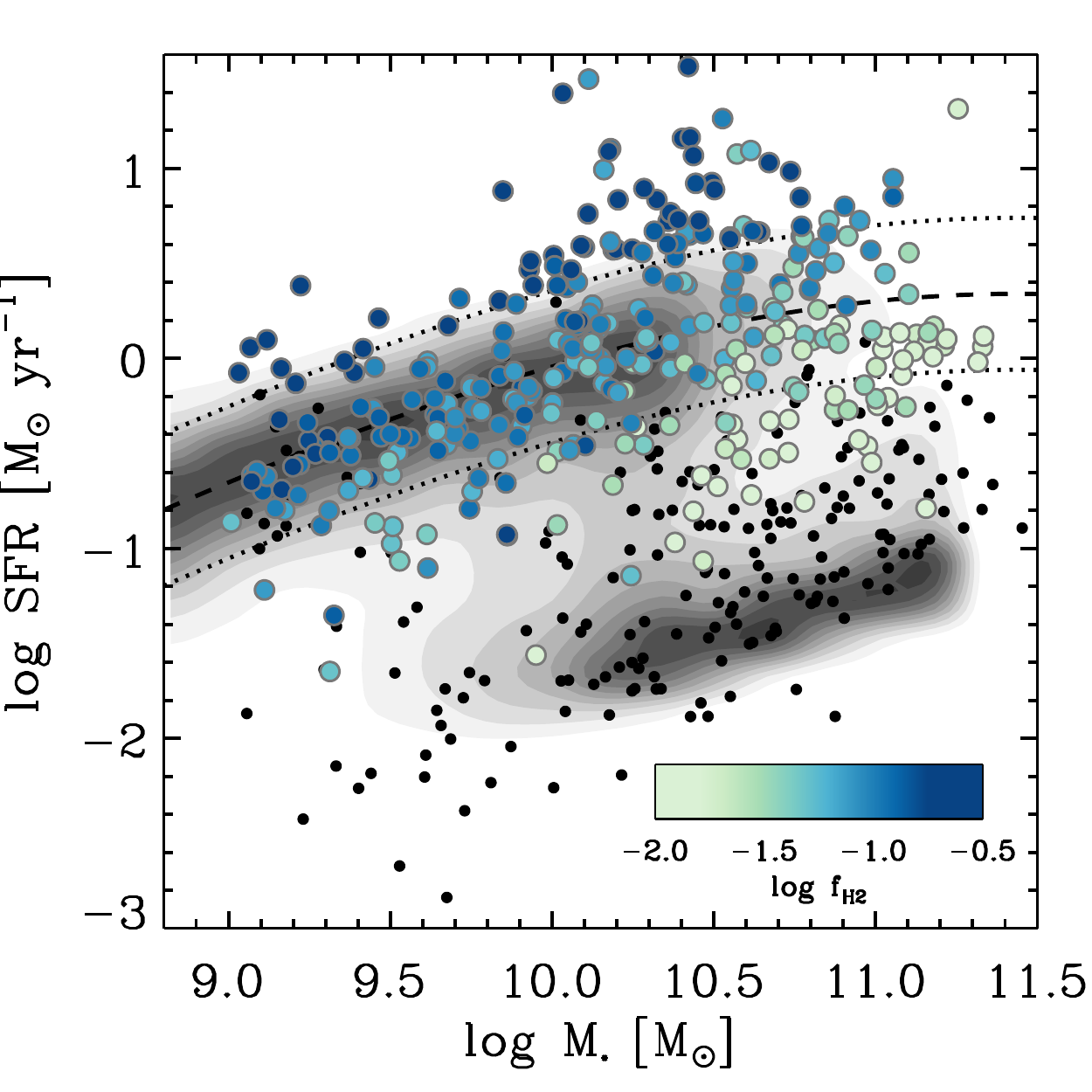}
\caption{Distribution of the xCOLD GASS sample in the \ms\ plane, color-coded by atomic gas mass fraction (left) and molecular gas mass fraction (right). The smaller black symbols are galaxies un-detected in the HI and CO(1-0) line, respectively. The grayscale contours show the overall SDSS population. The dashed and dotted lines indicate the position of the main sequence and the $\pm0.4$dex scatter around this relation, respectively. \label{msfig}}
\end{figure*}

\begin{figure}
\epsscale{1.1}
\plotone{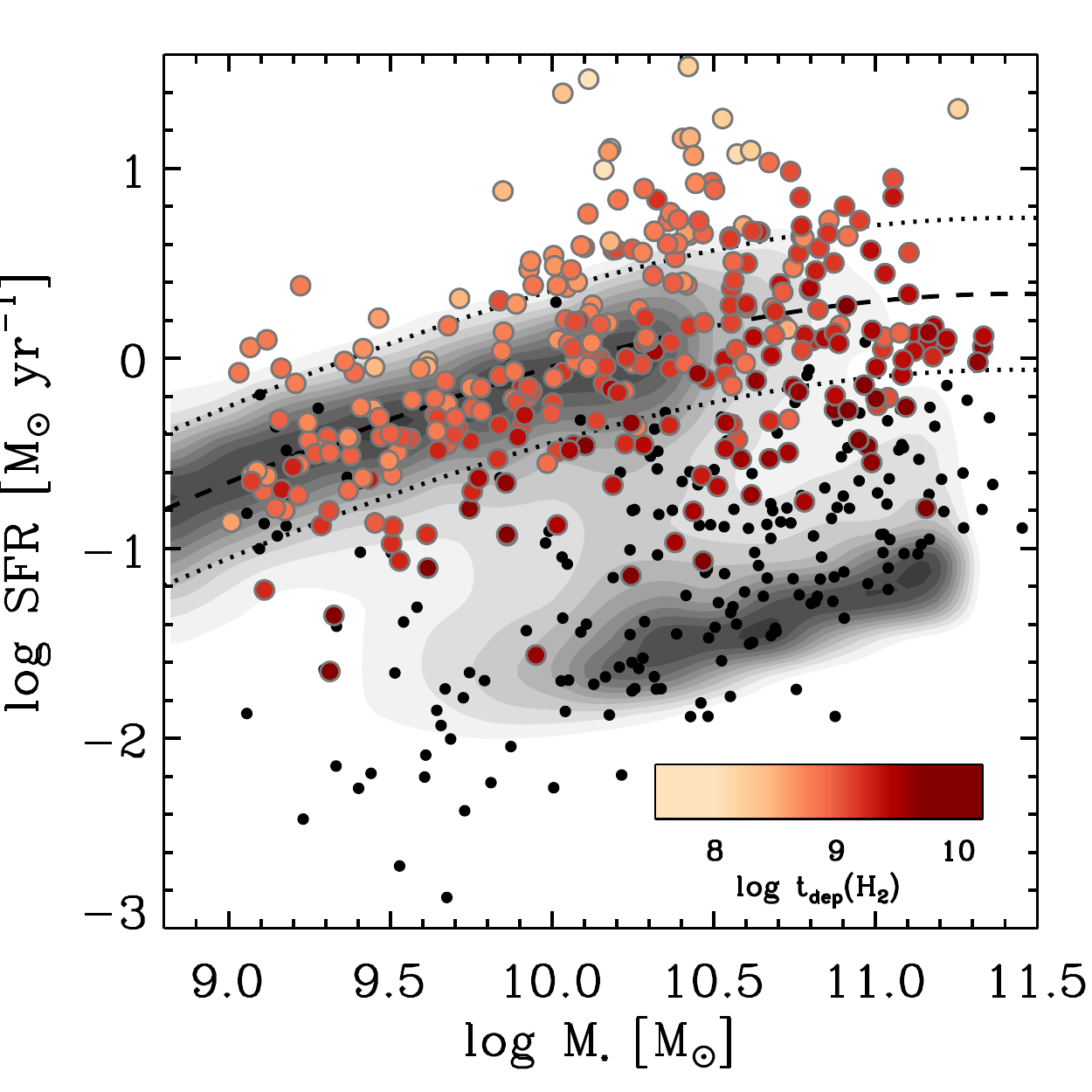}
\caption{Distribution of the xCOLD GASS sample in the \ms\ plane, color-coded by molecular gas depletion timescale. Symbols and lines as described in Fig. \ref{msfig}. \label{msfigtdep}}
\end{figure}

The distribution of the xCOLD GASS sample in the \ms\ plane is presented in Figure \ref{msfig}. As expected, the detection rate of the CO line goes from nearly 100\% for galaxies on and above the main sequence to zero for passive galaxies with very low star formation rates.  While individually un-detected, using a spectral stacking technique we measure in \citet{saintonge16} a mean molecular gas to stellar mass ratio of 0.6\% for the massive passive galaxies (\mstar$>10^{10}$\msun, $-12<\log {\rm SSFR}<-11$). {\edit Figures \ref{msfig} and \ref{msfigtdep}} also show how both molecular gas fraction (\fgas$\equiv$\mh$/$\mstar) and depletion time (\tdep$\equiv$\mh$/$SFR) vary systematically in the \ms\ plane. Galaxies above the main sequence have both higher gas fractions and shorter depletion times, while the reverse is true of below main sequence galaxies. As pointed out already in \citet{saintonge16}, it is the combination of variations in both gas contents and star formation efficiency that explains the position of galaxies in the \ms\ plane. Figures \ref{msfig} and \ref{msfigtdep} show that this trend extends to \mstar$<10^{10}$\msun.

\begin{figure*}
\epsscale{1.2}
\plotone{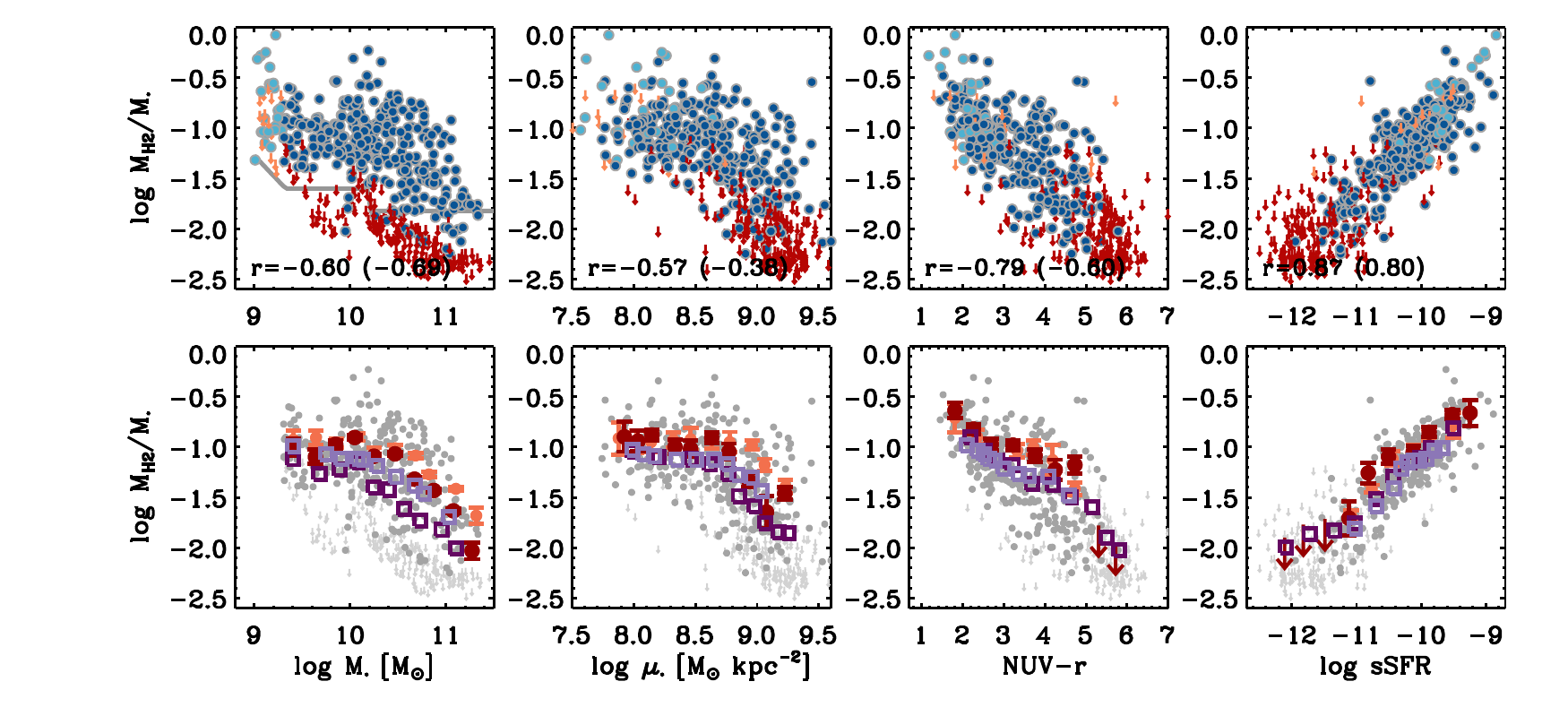}
\caption{Molecular gas mass fraction as a function of stellar mass, stellar mass surface density, NUV-r colour, and specific star formation rate. {\it Top row:} All xCOLD GASS galaxies, both with CO detections (circles) and upper-limits (arrows). Galaxies with $\log M_{\ast}<9.3$ where the upper limits are less constraining are shown in pale blue and orange, higher mass galaxies in dark blue and red. The correlation coefficient between \fgas\ and the x-axis parameter for the CO detections is given in each panel. {\it Bottom row:} Mean \fgas\ scaling relations obtained by stacking of all xCOLD GASS galaxies are shown with red circles, those for main sequence galaxies only ($|$\deltams$|<0.4$) as orange circles. The open squares show the weighted medians of $\log$\fgas\ in bins of the x-axis parameter for all xCOLD GASS and main sequence only in dark and pale purple, respectively.    \label{fH2data}}
\end{figure*}

To explore further the dependence of gas fractions on integrated galaxy properties, scaling relations for \fgas\ are presented in Figure \ref{fH2data}. The relation between \fgas\ and \mstar\ shows a mild dependence at \mstar$<10^{10.5}$\mstar\ but a sharp drop off afterwards. This is in contrast with the atomic gas fractions, \fhi, which increase steadily as \mstar\ decreases. Indeed, while the mean molecular-to-atomic ratio for massive galaxies is $<R_{mol}>\sim0.35$, it drops to only 0.1 for the lowest mass galaxies in the sample (see also Catinella et al. 2017). The dependence of both \fgas\ {\edit and the molecular ratio} (\rmol$\equiv M_{H_2}/M_{HI}$) on \mstar\ in the xCOLD GASS sample is qualitatively similar to those observed previously in the $10^9-10^{11.5}$\msun\ stellar mass range by combining the COLD GASS data with CO observations from the Herschel Reference Survey \citep{boselli14} and from ALLSMOG \citep{bothwell14,cicone17}. However, the xCOLD GASS scaling relations have the advantage of being based on a sample that is larger, deeper and more homogeneous. The dependence of \fgas\ on morphology (as parametrised here by the stellar mass surface density, \must) is similar, with gas fractions decreasing sharply when entering the regime of bulge-dominated galaxies ($\log \mu_{\ast}\sim8.7$).

The strongest correlations however are between \fgas\ and quantities relating to SFR. While \nuvr\ is a proxy for SSFR since NUV colour traces ongoing star formation and $r$-band the older stellar population, the correlation with \fgas\ is not as strong as with SSFR itself. The main difference between \nuvr\ and SSFR is that the former does not consider dust obscuration, which explains the weaker correlation as molecular gas is known to be best linked with dust-obscured star formation as measured in the mid- or far-infrared \citep[e.g.][]{leroy08}.

To best compare these results with other studies and the results of models and simulations we provide in Table \ref{fh2relations}  \fgas\ scaling relations measured in two different ways. The first method relies on a spectral stacking technique \citep{fabello11a,saintonge12,saintonge16} whose main advantage is to allow CO detections and non-detections to be combined fairly. The disadvantage is that the stacking process is linear by nature, whereas the distribution of \fgas\ values in selected bins tends to be lognormal. The second method we use is to calculate weighted median values of $\log$~\fgas\ {\edit (we refer to this as the ``binning method" in Table \ref{fh2relations})}. This technique has the opposite pros and cons: it makes the better assumption that the distribution of \fgas\ values in selected bins is lognormal, but has to assume for the CO non-detections the value of \fgas\ set by the upper limit. The latter is not a problem if there are few non-detections in any given bin and if the upper limits are well separated from the detections, but becomes an issue when the rate of non-detections is significant. For both averaging methods, we also provide the \fgas\ scaling relations for the complete xCOLD GASS sample, as well as for main-sequence galaxies only ($|$\deltams$|<0.4$); readers are cautioned to select with care the optimal set of \fgas\ scaling relations for their specific purposes. 

For both methods, the errors reported include the statistical errors associated with the IRAM calibration and aperture corrections \citep[which taken together are typically $20\%$ for a given galaxy, see][]{COLDGASS1}, and the sampling error determined from bootstrapping. This does not however include the systematic error on the conversion function, \xco, which \citet{accurso17b} estimate to be 35\%.   

Throughout this work, we adopt the CO-to-H$_2$ conversion function of \citet{accurso17b}, which is primarily a function of metallicity, with a secondary dependence on the offset of the galaxy from the star formation main sequence. In that work (see their Figs. 8 \& 9), we show that were we to use a constant Galactic conversion factor across the xCOLD GASS sample, the scatter around key scaling relations for \fgas\ and the depletion timescale is increased. \citet{tacconi17} also show the impact of using different prescription for the  CO-to-H$_2$ conversion function on molecular gas scaling relations, combining xCOLD GASS with sample of high redshift galaxies. 

\begin{deluxetable*}{ccccccccccccccc}
\tabletypesize{\footnotesize}
\tablecolumns{15}
\tablecaption{\fgas\ scaling relations \label{fh2relations}}
\tablehead{
\multicolumn{7}{c}{All xCOLD GASS} & \colhead{} & \multicolumn{7}{c}{Main sequence only} \\ 
\cline{1-7} \cline{9-15}
\multicolumn{3}{c}{Stacking} & \colhead{} & \multicolumn{3}{c}{Binning} & \colhead{} & \multicolumn{3}{c}{Stacking} & \colhead{} & \multicolumn{3}{c}{Binning} \\
\cline{1-3} \cline{5-7} \cline{9-12} \cline{13-15}
\colhead{$<x>$} & \colhead{N} & \colhead{$\log<$\fgas$>$} &  & \colhead{$<x>$} & \colhead{N} & \colhead{$<\log$\fgas$>$} & &  \colhead{$<x>$} & \colhead{N} & \colhead{$\log<$\fgas$>$} &  & \colhead{$<x>$} & \colhead{N} & \colhead{$<\log$\fgas$>$} } 
\startdata
\cutinhead{$x=\log$\mstar}
9.407 &       36 & $-1.01\pm0.041$ & & 9.388 &       48 & $-1.11\pm0.058$ & & 9.406 &       23 & $-0.93\pm0.049$ & & 9.366 &       18 & $-0.99\pm0.072$ \\
9.638 &       34 & $-1.08\pm0.035$ & & 9.669 &       48 & $-1.28\pm0.078$ & & 9.643 &       19 & $-0.89\pm0.033$ & & 9.465 &       18 & $-1.05\pm0.068$ \\
9.848 &       41 & $-0.97\pm0.020$ & & 9.915 &       48 & $-1.23\pm0.088$ & & 9.854 &       24 & $-0.96\pm0.023$ & & 9.647 &       18 & $-1.12\pm0.077$ \\
10.04 &       61 & $-0.90\pm0.011$ & & 10.10 &       48 & $-1.16\pm0.080$ & & 10.07 &       26 & $-0.89\pm0.017$ & & 9.882 &       18 & $-1.10\pm0.066$ \\
10.24 &       66 & $-1.08\pm0.013$ & & 10.25 &       48 & $-1.39\pm0.091$ & & 10.25 &       22 & $-1.05\pm0.023$ & & 10.01 &       18 & $-1.11\pm0.079$ \\
10.46 &       70 & $-1.05\pm0.009$ & & 10.41 &       48 & $-1.40\pm0.124$ & & 10.46 &       16 & $-0.99\pm0.018$ & & 10.14 &       18 & $-1.20\pm0.074$ \\
10.67 &       63 & $-1.34\pm0.011$ & & 10.58 &       48 & $-1.62\pm0.087$ & & 10.68 &       20 & $-1.10\pm0.014$ & & 10.37 &       18 & $-1.27\pm0.088$ \\
10.87 &       56 & $-1.41\pm0.009$ & & 10.75 &       48 & $-1.75\pm0.086$ & & 10.82 &       21 & $-1.26\pm0.013$ & & 10.56 &       18 & $-1.39\pm0.114$ \\
11.07 &       48 & $-1.66\pm0.012$ & & 10.90 &       48 & $-1.83\pm0.088$ & & 11.10 &       17 & $-1.41\pm0.010$ & & 10.77 &       18 & $-1.44\pm0.108$ \\
11.27 &       15 & $-2.02\pm0.041$ & & 11.08 &       48 & $-2.01\pm0.075$ & & 11.31 &        5 & $-1.68\pm0.039$ & & 11.12 &       18 & $-1.70\pm0.128$ \\
\cutinhead{$x=\log \mu_{\ast}$}
7.876 &        9 & $-0.90\pm0.076$ & & 8.025 &       42 & $-1.03\pm0.049$ & & 7.876 &        7 & $-0.91\pm0.079$ & & 7.976 &       17 & $-1.04\pm0.093$ \\
8.032 &       24 & $-0.94\pm0.049$ & & 8.211 &       42 & $-1.09\pm0.068$ & & 8.025 &       20 & $-0.94\pm0.055$ & & 8.117 &       17 & $-1.05\pm0.068$ \\
8.146 &       35 & $-0.87\pm0.027$ & & 8.427 &       42 & $-1.10\pm0.058$ & & 8.139 &       19 & $-0.95\pm0.054$ & & 8.257 &       17 & $-1.09\pm0.072$ \\
8.333 &       29 & $-1.00\pm0.035$ & & 8.589 &       42 & $-1.16\pm0.097$ & & 8.320 &       16 & $-0.91\pm0.038$ & & 8.431 &       17 & $-1.11\pm0.074$ \\
8.459 &       39 & $-0.99\pm0.039$ & & 8.709 &       42 & $-1.26\pm0.114$ & & 8.460 &       20 & $-0.93\pm0.057$ & & 8.509 &       17 & $-1.13\pm0.071$ \\
8.636 &       52 & $-0.91\pm0.027$ & & 8.850 &       42 & $-1.49\pm0.124$ & & 8.631 &       24 & $-0.92\pm0.041$ & & 8.682 &       17 & $-1.12\pm0.089$ \\
8.777 &       48 & $-1.04\pm0.042$ & & 8.962 &       42 & $-1.60\pm0.080$ & & 8.777 &       28 & $-0.94\pm0.051$ & & 8.764 &       17 & $-1.15\pm0.084$ \\
8.931 &       65 & $-1.34\pm0.036$ & & 9.044 &       42 & $-1.68\pm0.113$ & & 8.958 &       22 & $-0.97\pm0.022$ & & 8.810 &       17 & $-1.27\pm0.162$ \\
9.076 &       60 & $-1.64\pm0.079$ & & 9.135 &       42 & $-1.84\pm0.062$ & & 9.060 &       12 & $-1.18\pm0.030$ & & 8.956 &       17 & $-1.32\pm0.110$ \\
9.223 &       74 & $-1.46\pm0.030$ & & 9.229 &       42 & $-1.84\pm0.092$ & & 9.230 &       15 & $-1.39\pm0.032$ & & 9.060 &       17 & $-1.42\pm0.121$ \\
\cutinhead{$x=$\nuvr}
\nodata &        0 & \nodata & & 2.140 &       46 & $-0.90\pm0.057$ & & \nodata &        0 & \nodata & & 2.186 &       18 & $-0.99\pm0.078$ \\
1.807 &       16 & $-0.64\pm0.042$ & & 2.510 &       46 & $-1.03\pm0.049$ & & 1.809 &        5 & $-0.71\pm0.069$ & & 2.337 &       18 & $-1.04\pm0.082$ \\
2.260 &       54 & $-0.82\pm0.025$ & & 2.730 &       46 & $-1.13\pm0.054$ & & 2.347 &       36 & $-0.89\pm0.036$ & & 2.455 &       18 & $-1.07\pm0.083$ \\
2.690 &       72 & $-0.96\pm0.025$ & & 3.066 &       46 & $-1.18\pm0.064$ & & 2.690 &       57 & $-0.99\pm0.031$ & & 2.550 &       18 & $-1.12\pm0.070$ \\
3.220 &       54 & $-0.97\pm0.023$ & & 3.570 &       46 & $-1.36\pm0.073$ & & 3.320 &       35 & $-1.02\pm0.033$ & & 2.700 &       18 & $-1.13\pm0.060$ \\
3.760 &       54 & $-1.09\pm0.031$ & & 3.910 &       46 & $-1.38\pm0.092$ & & 3.720 &       25 & $-0.98\pm0.033$ & & 2.970 &       18 & $-1.20\pm0.094$ \\
4.250 &       44 & $-1.22\pm0.050$ & & 4.530 &       46 & $-1.50\pm0.094$ & & 4.186 &       18 & $-1.15\pm0.081$ & & 3.340 &       18 & $-1.28\pm0.094$ \\
4.730 &       47 & $-1.16\pm0.041$ & & 5.120 &       46 & $-1.60\pm0.113$ & & 4.730 &       13 & $-1.42\pm0.031$ & & 3.500 &       18 & $-1.24\pm0.095$ \\
5.305 &       56 & $-1.78\pm-9.00$ & & 5.460 &       46 & $-1.90\pm0.070$ & & \nodata &        2 & \nodata & & 4.006 &       18 & $-1.29\pm0.089$ \\
5.720 &       68 & $-1.95\pm-9.00$ & & 5.790 &       46 & $-2.03\pm0.049$ & & \nodata &        0 & \nodata & & 4.450 &       18 & $-1.45\pm0.122$ \\
\cutinhead{$x=\log$SSFR}
-12.1 &       39 & $-1.91\pm-9.00$ & & -12.1 &       47 & $-1.99\pm0.064$ & & \nodata &        0 & \nodata & & -10.9 &       18 & $-1.82\pm0.076$ \\
-11.8 &       47 & $-1.76\pm-9.00$ & & -11.8 &       47 & $-1.88\pm0.087$ & & \nodata &        0 & \nodata & & -10.7 &       18 & $-1.62\pm0.106$ \\
-11.4 &       54 & $-1.71\pm-9.00$ & & -11.4 &       47 & $-1.81\pm0.089$ & & \nodata &        0 & \nodata & & -10.5 &       18 & $-1.41\pm0.095$ \\
-11.1 &       49 & $-1.70\pm0.084$ & & -11.1 &       47 & $-1.76\pm0.075$ & & -11.0 &       13 & $-1.68\pm0.026$ & & -10.3 &       18 & $-1.23\pm0.085$ \\
-10.8 &       44 & $-1.25\pm0.049$ & & -10.7 &       47 & $-1.50\pm0.117$ & & -10.7 &       19 & $-1.41\pm0.019$ & & -10.2 &       18 & $-1.15\pm0.074$ \\
-10.5 &       52 & $-1.09\pm0.034$ & & -10.5 &       47 & $-1.30\pm0.062$ & & -10.4 &       33 & $-1.13\pm0.035$ & & -10.1 &       18 & $-1.14\pm0.063$ \\
-10.1 &       80 & $-1.04\pm0.029$ & & -10.2 &       47 & $-1.15\pm0.047$ & & -10.1 &       73 & $-1.03\pm0.029$ & & -10.0 &       18 & $-1.11\pm0.075$ \\
-9.87 &       65 & $-0.85\pm0.024$ & & -10.0 &       47 & $-1.10\pm0.052$ & & -9.88 &       46 & $-0.88\pm0.029$ & & -9.96 &       18 & $-1.14\pm0.074$ \\
-9.51 &       38 & $-0.68\pm0.023$ & & -9.79 &       47 & $-1.01\pm0.047$ & & -9.50 &        8 & $-0.77\pm0.062$ & & -9.87 &       18 & $-1.04\pm0.076$ \\
-9.24 &        7 & $-0.66\pm0.064$ & & -9.49 &       47 & $-0.81\pm0.057$ & & \nodata &        0 & \nodata & & -9.66 &       18 & $-1.00\pm0.070$ 
\enddata
\end{deluxetable*}

\subsection{Depletion time scaling relations}

In previous studies, we have reported systematic variations of the molecular gas depletion time, \tdep$=$\mh$/$SFR, across the galaxy populations \citep{COLDGASS2,saintonge12,saintonge16,genzel15,tacconi17}. These variations have been associated with dynamical effects that can increase or decrease the pressure within the ISM thus influencing how much of the cold molecular gas can reach the high densities of pre-stellar cores. The story is different for the atomic gas depletion time,  \tdepHI$=$\mhi$/$SFR, which has been measured to have much larger galaxy-to-galaxy variations, but no systematic dependence on global galaxy properties \citep{GASS2}. Here we revisit these results, but now expanding the analysis to \mstar$<10^{10}$\msun, as made possible by xCOLD GASS and xGASS. 

\begin{figure*}
\epsscale{1.2}
\plotone{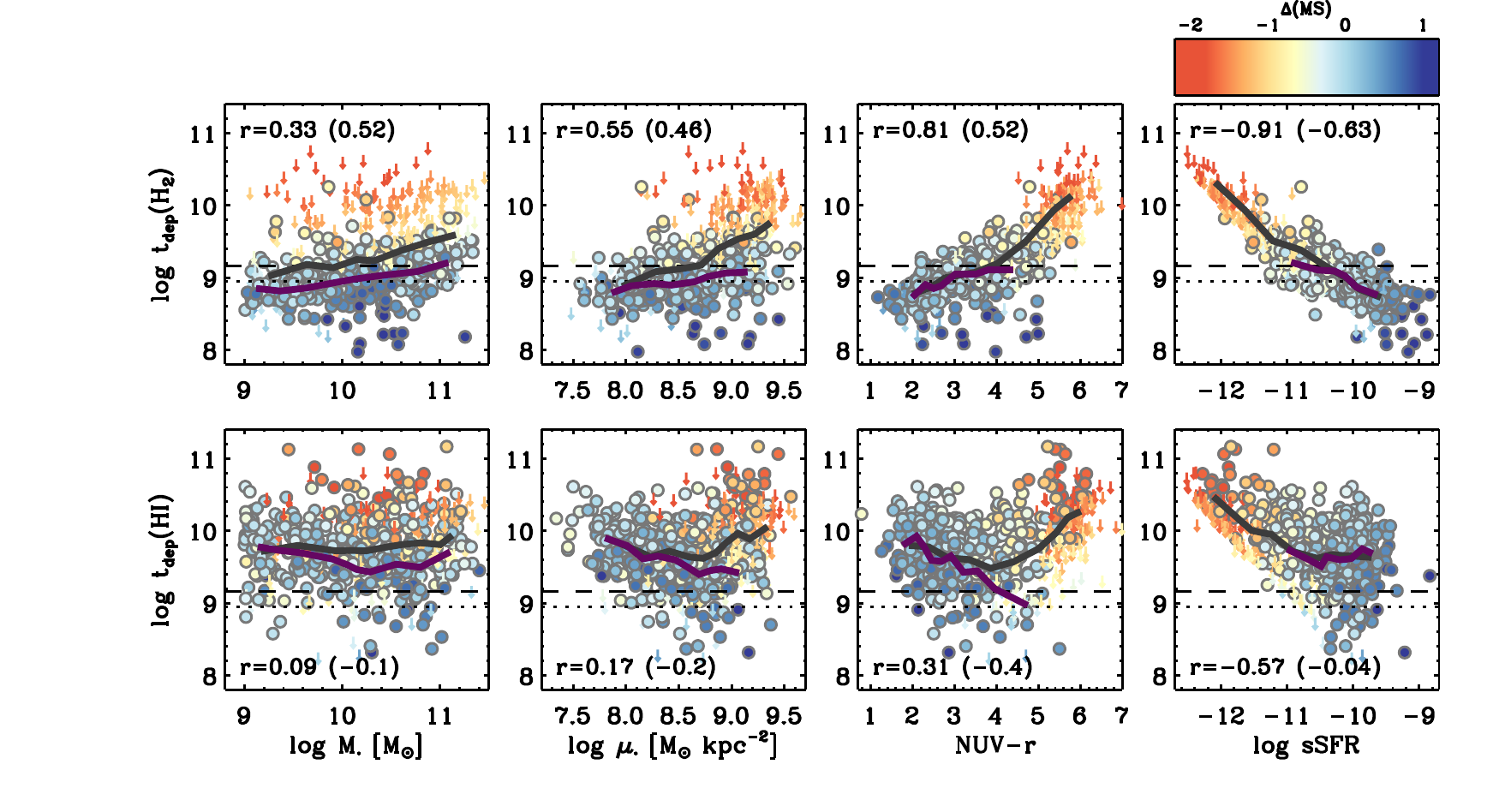}
\caption{Comparison between depletion timescales for molecular (top row) and atomic gas (bottom row), and \mstar, stellar mass surface density, \nuvr\ and SSFR.  Galaxies are color-coded by their offset from the main sequence, \deltams. Weighted medians in bins of the x-axis parameter are given including all galaxies in the xCOLD GASS / xGASS samples. In each panel, the correlation coefficient is given for all galaxies (and main sequence galaxies only), while the dashed line shows the weighted median value of $\log$\tdep\ across the entire sample and the dotted line is the median value for main sequence galaxies only.   \label{tdepHIH2}}
\end{figure*}

Figure \ref{tdepHIH2} shows how both \tdep\ and \tdepHI\ vary as a function of integrated properties. It confirms previous observations that \tdepHI\ is independent of \mstar, and at fixed mass has larger scatter than \tdep. When looking at main sequence galaxies, the weighted mean values and scatter are $<\log$~\tdep$>=8.98\pm0.27$ and  $<\log$~\tdepHI$>=9.65\pm0.44$. The longest molecular gas depletion times we measure are $<10^{10}$~yr, while atomic gas depletion times can be in excess of $10^{11}$~yr. This illustrates clearly how galaxies may have large reservoirs of cold atomic gas which are not associated with the star formation process, as may be the case for example in early-type galaxies \citep{serra11,serra14,gereb16}, and in low mass galaxies.

\subsection{Molecular gas in AGN hosts} 

\begin{figure}
\epsscale{1.1}
\plotone{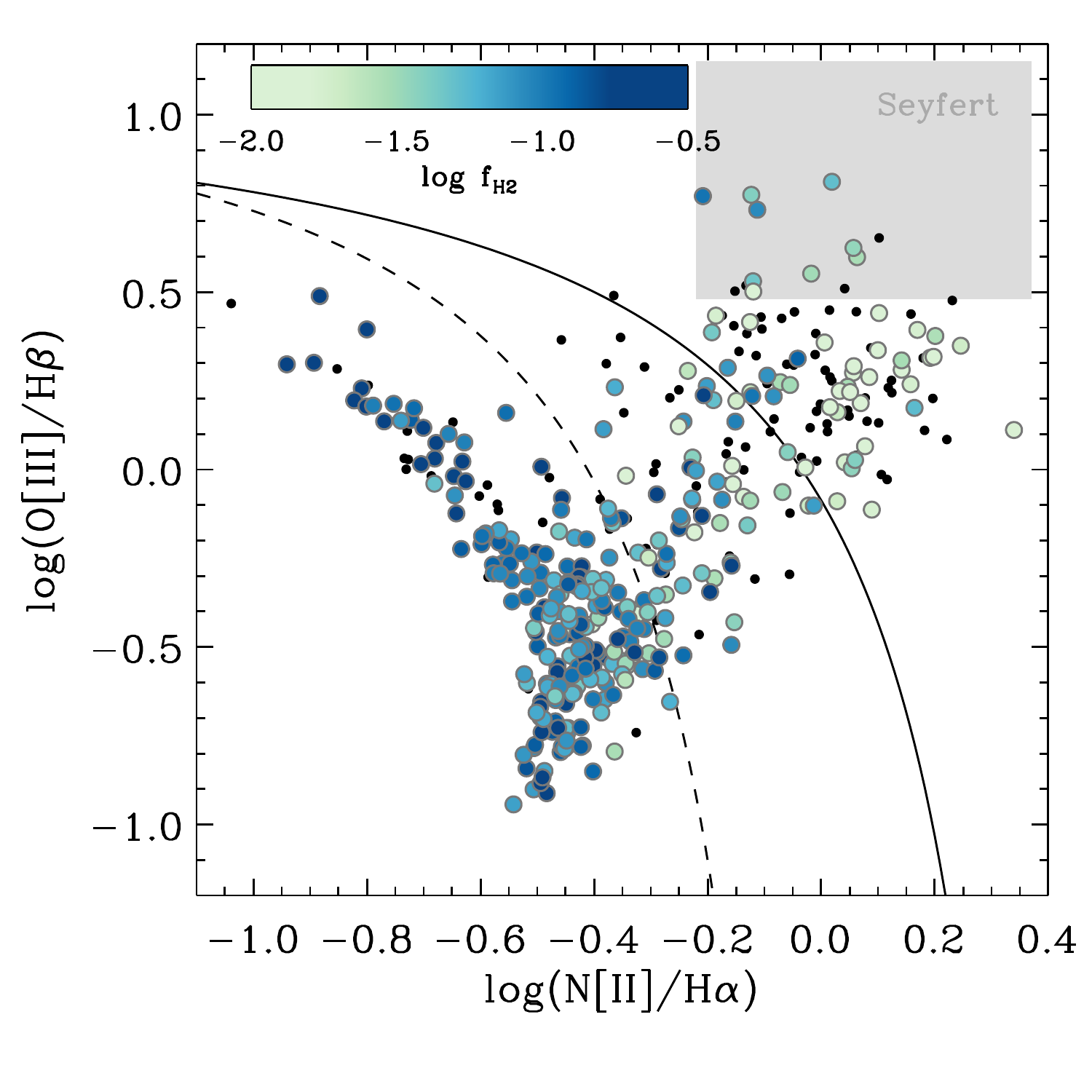}
\caption{Distribution of the xCOLD GASS galaxies in the BPT diagram. Shown are the galaxies with S/N$>3$ in the SDSS spectrum in all four emission lines.  Galaxies are color-coded by their molecular gas mass fractions, with CO non-detections shown as the smaller black symbols. The empirical relation of \citet{kauffmann03AGN} and the theoretical relation of \citet{kewley01} to separate systems where gas is ionised by star-formation and AGN activity are shown as the dashed and solid lines, respectively. The grey shaded regions shows the locus of Seyfert galaxies.   \label{bptfig}}
\end{figure}

Gas offers a possible explanation for the observed correlation between supermassive black hole mass and the properties of its host galaxy \citep[e.g. bulge mass and stellar velocity dispersion,][]{magorrian98,haring04,merloni10}. The correlation can be expected under the assumption that both host galaxy and supermassive black hole grow together through merging and as a function of the availability of gas. Evidence for this fundamental link between black hole and galaxy growth includes the remarkable similarity between the redshift evolution of star formation activity (i.e the Lilly-Madau diagram) and that of the rate of accretion onto supermassive black holes \citep[e.g.][]{boyle98,aird10,merloni13}. Under this scenario, one would expect a {\it positive} correlation between the molecular gas fraction of galaxies and incidence of AGN activity. 

On the other hand, the ability of AGN to drive massive molecular gas outflows \citep{sturm11,costa14,cicone14} suggests a different scenario for the link between molecular gas and AGN activity. Feedback from AGN is an appealing quenching mechanism, as it can explain why most quenched galaxies are bulge-dominated \citep{somerville08,bell08}. If the AGN-driven molecular outflows are indeed an important quenching agent, then we would expect a {\it negative} correlation between AGN activity and the molecular gas contents of galaxies. Whether the observations are consistent with either of these scenarios depends not only on the physical mechanisms at play, but perhaps mostly on the timescales involved in the various transformations. 

What is then the observed link between the molecular gas contents of galaxies and AGN activity? In Figure \ref{bptfig} we show the distribution of the xCOLD GASS sample in the BPT diagram \citep{baldwin81}, with the color-coding representing the molecular gas mass fraction, $f_{H2}$. Visual inspection suggests that galaxies in the AGN part of the BPT diagram tend to be gas-poor, with low gas fractions and a higher rate of non-detection, but gas fractions possibly increasing back in Seyfert galaxies, especially those with the highest [OIII]/H$_{\beta}$ line ratios. However, to properly assess whether active and inactive galaxies differ in terms of their molecular gas contents, we must control for SSFR, given the strong correlation with \fgas (see Fig. \ref{fH2data}). After extracting samples of active and inactive galaxies matched in SSFR from the high mass COLD GASS sample (\mstar$>10^{10}$\msun), the KS statistics reveals with a test probability of 0.12 that BPT-selected AGN hosts have gas fractions slightly lower than matched xCOLD GASS inactive galaxies (1.4\% and 2.1\% respectively). The test was done considering both CO detections and non-detection. However, if focussing only on the Seyfert sub-sample, the test reveals that they are indistinguishable from the matched inactive control sample in terms of molecular gas fraction, though with only a handful of Seyferts in the xCOLD GASS sample, the statistics are poor. 

The slightly below-average molecular gas contents of AGN hosts is consistent with the analysis of \citet{shimizu15}, which showed that the AGN in COLD GASS and in the {\it Swift}/BAT sample tend to lie below the main sequence, where we have shown gas fractions to decrease (Fig. \ref{msfig}). The analysis of \citet{shimizu15} also reveals that, on the other hand, the brightest AGN tend to be in high mass star-forming galaxies with a high merger rate \citep[see also ][]{koss11}. While the xCOLD GASS sample only contains a handful of these bright AGN, they do appear to have higher gas fractions than weaker AGN and on par with SSFR-matched inactive galaxies. 

By targeting more nearby moderately-luminous Seyferts from the LLAMA survey with APEX and JCMT, Rosario et al. (2017, submitted) were able to study the possible impact of AGN on molecular gas in the central regions, where the influence of the AGN (if any) should be greatest. They report central gas fractions and star formation efficiencies that are similar between active and matched control inactive galaxies, which suggests that nuclear radiation does not couple efficiently with the surrounding molecular gas. \citet{kirkpatrick14} also find no impact of AGN on star formation efficiency in galaxies from the 5MUSES sample. 

Coming back to the two scenarios described at the beginning of this section, the xCOLD GASS results tentatively suggest that the correlation between molecular gas and AGN activity is a function of AGN strength; normal or elevated molecular gas contents sustain fuelling onto Seyfert nuclei, while the hosts of weaker AGN show depletion of their molecular gas reservoirs, possibly after a period of more intense activity and feedback. Observations of molecular line emission in large, complete samples of galaxies hosting AGN with a wide range of luminosities will be key to further disentangle the competing effects of fuelling and feedback.

\section{Discussion}

\subsection{From circumgalactic gas to star forming regions}

There is significant diagnostic power in the cold gas contents of galaxies. For example, identifying whether galaxies are quiescent because they are gas-poor, or because they are very inefficient at converting any cold gas they may have into stars, leads to vastly different interpretations as to the mechanisms responsible for quenching. The xCOLD GASS view on this is that the availability of molecular gas is mostly responsible for determining the star formation rate of a galaxy (\fgas\ correlates strongly with SSFR; Fig. \ref{fH2data}). However, the data show that the depletion timescale of the molecular gas also correlates with star formation activity (Fig. \ref{tdepHIH2}). The COLD GASS-low sample allows us here to extend the conclusions of \citet{saintonge16} that, on average, the SSFR of galaxies is set by both the molecular gas fraction and the star formation efficiency. 

Having established this, the questions now become: (1) what sets the molecular gas fraction of a galaxy? and (2) why does the molecular gas depletion time vary across the local galaxy population?  To address these questions, we need to look on scales much larger and much smaller than the molecular discs of the xCOLD GASS galaxies. 

For the first question, it is helpful to consider the broader galactic ecosystem. Using the Cosmic Origins Spectrograph on HST, the circumgalactic medium (CGM) of galaxies from GASS was probed via absorption features in the spectra of background quasars.  \citet{borthakur15} report a significant correlation between Ly$\alpha$ equivalent width in the CGM and \fhi\ in the ISM. This correlation is even stronger than the correlation between Ly$\alpha$ equivalent width and the SSFR, a result consistent with our observation in Figure \ref{tdepHIH2} that HI is less tightly connected to star formation than H$_2$. This implies a physical connection between the atomic gas in the ISM of galaxies and their CGM, bridging the significant scale difference between the two components, and consistent with a picture where the cold atomic gas reservoir of galaxies is fed by accretion of gas from the CGM \citep{borthakur15}. 

The next step in the journey of gas from CGM to stars is the conversion of atomic gas to the molecular phase. How the molecular ratio, $R_{mol}\equiv M_{H_2}/M_{HI}$, varies across the xCOLD GASS sample is studied in detail in Catinella et al. (2017), but as can be seen in Figures \ref{tdepHIH2} and \ref{modcomp}, the global atomic and molecular gas masses of galaxies relate differently with SFR and \mstar\ across the sample. For example, the highest values of $R_{mol}$ are observed in galaxies with high stellar masses, particularly at high SFRs \citep{saintonge16}, and while $R_{mol}\sim0.3$ in galaxies with \mstar$>10^{10}$\mstar\ \citep{COLDGASS1}, the ratio plummets to a mean value of $\sim0.1$ in galaxies with \mstar$<10^{10}$\mstar\ (Catinella et al. 2017). 

Forming molecular gas clouds with T$\sim20$K and densities of $\sim10^3$cm$^{-3}$, the conditions of the gas probed by our CO(1-0) observations, appears to be a necessary but not sufficient condition for star formation to efficiently proceed. Molecular gas traced by CO(1-0) may be too diffuse to lead to star formation, and we have previously hypothesised that ``contamination" from this diffuse molecular gas may be the reason we observe variations in \tdep\ across the COLD GASS sample \citep{COLDGASS2}, based on several studies which have shown how SFR correlates more tightly and linearly with the luminosity of molecular lines with higher critical densities than CO(1-0), such as HCN(1-0) \citep[e.g.][]{gao04,gracia06,garcia12}. While these pioneering studies mostly targeted intensely star forming, infrared-bright galaxies, recent instrumentation improvements have allowed two important breakthroughs: the observation of molecular lines such as HCN and HCO$^{+}$(1-0) in main-sequence star-forming galaxies and even early-type galaxies \citep{crocker12}, and the mapping of these lines at kiloparsec scales rather than integrated over entire galaxies. The latter is particularly important to link the global, statistical studies with high resolution observations of individual clouds in the Milky Way and other very nearby galaxies \citep{wu05,buchbender13}. 

A very exciting recent turn of events in the field of star formation studies is the realisation that the systematic variations in star formation efficiency across the galaxy population first revealed by COLD GASS trickle down to small scales. For example, \citet{usero15} show that the dense gas fraction depends strongly on local conditions within the disk of a galaxy, and by focussing on M51, \citet{bigiel16} reveal that spatial variations in star formation efficiency and dense gas fraction are associated with the local environment, and in particular pressure conditions. Even more interestingly, \citet{hughes13,hughes16} show that it is the properties of GMCs themselves which vary not only as a function of local environment but with the global properties of galaxies (their total \mstar\ or gas surface density for example). 

The systematic observations of molecular gas throughout the local galaxy population with xCOLD GASS have helped to uncover this vast multi-scale relation between the CGM, to the atomic ISM, to the cold molecular gas, to the star formation process on small scales. This complex chain of correlations between the properties of the CGM down to star formation efficiency at the smallest of scales suggest that ``whole cloud" theories for star formation \citep[e.g.][]{krumholz05,federrath12} may have to also take into account the larger scale environment that regulates the availability of gas on small scales.

\subsection{The importance of cold gas scaling relations for galaxy formation simulations}

\begin{figure*}
\epsscale{1.1}
\plottwo{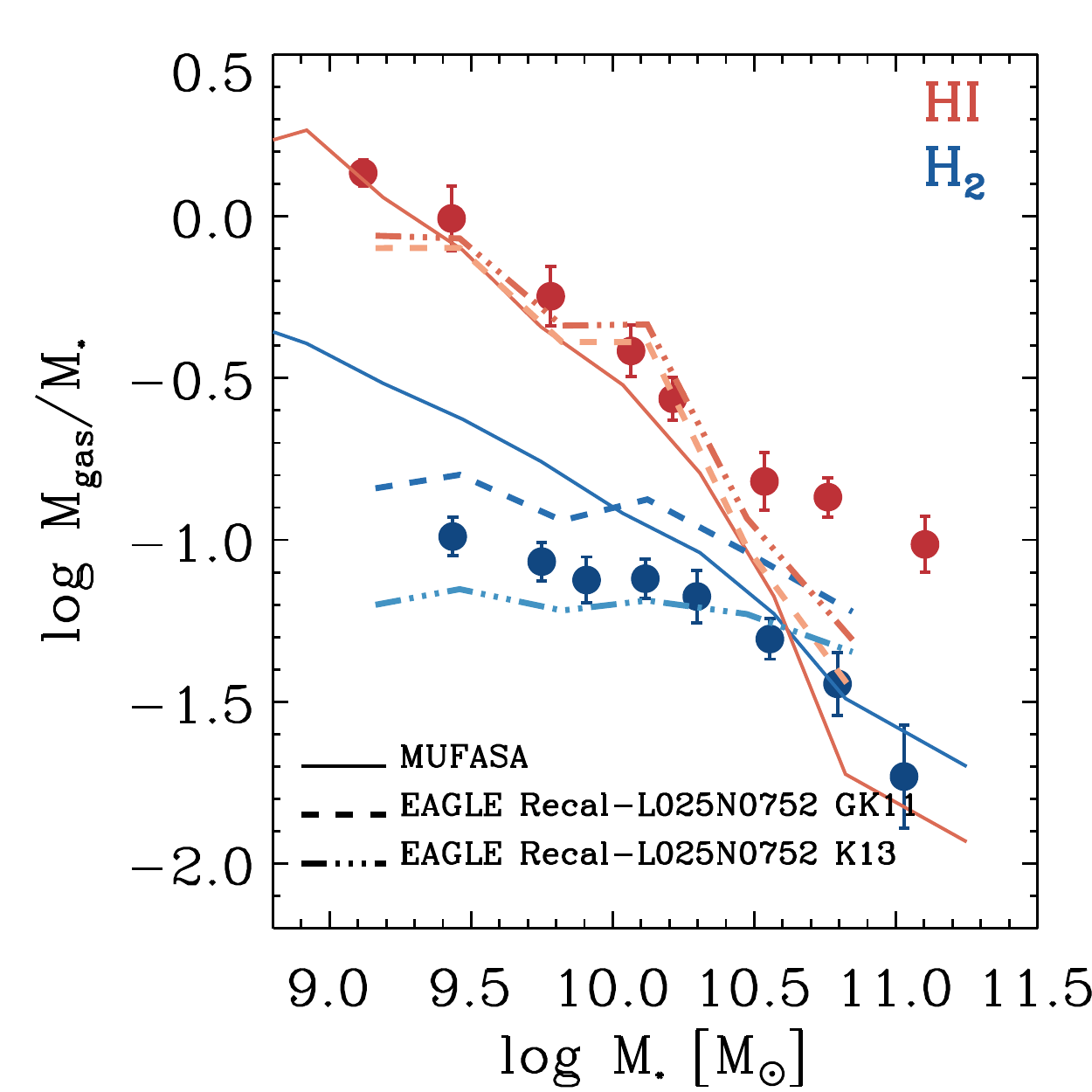}{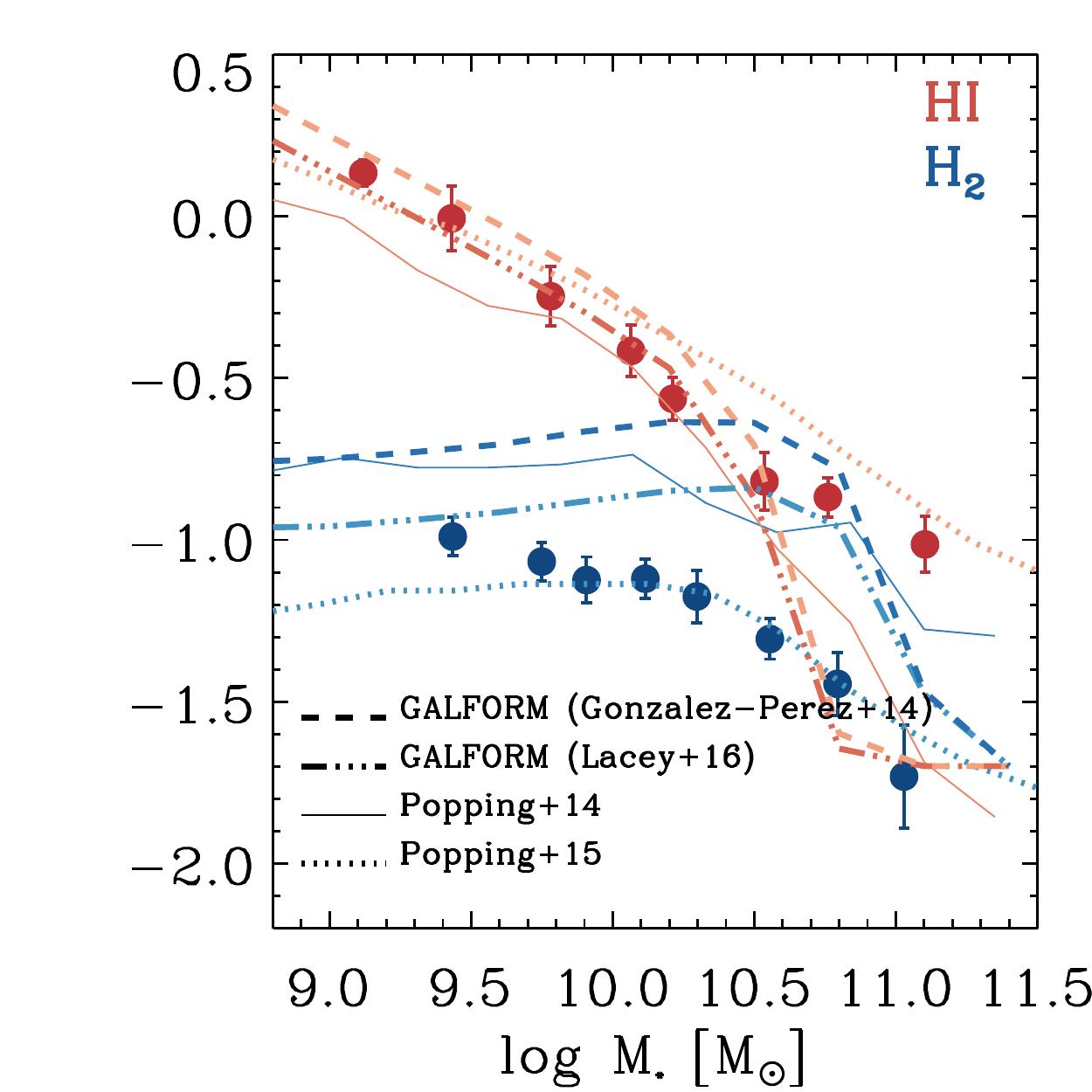}
\caption{The molecular and atomic gas fraction dependence on \mstar, from xGASS/xCOLD GASS, compared with model predictions. {\it Left:} comparison with the hydrodynamic simulations MUFASA \citep{dave17} and EAGLE \citep{lagos15}.  The two EAGLE models differ in the prescription used to predict how much of the total cold gas is molecular. {\it Right:} comparison with semi-analytical models, two different versions of GALFORM \citep{gonzalezperez14,lacey16}, and the work of \citet{popping14} and \citet{popping15}. \label{modcomp}}
\end{figure*}

The cold gas scaling relations also have strong constraining power for cosmological simulations. For example, since simulations and models typically use the empirical constraint of the $z=0$ stellar mass function to tune their parameters, it leaves the gas mass function as a powerful reference to post-facto assess their predictive power. To illustrate this, we show in Figure \ref{modcomp} how the xCOLD GASS scaling relations between molecular and atomic gas fractions and \mstar\ compare to the results of semi-analytical and hydrodynamical models.  The comparison is shown for star-forming galaxies. The most recent generation of large hydrodynamical simulations such as EAGLE, ILLUSTRIS and MUFASA can track the cold (T$<10^4$~K) ISM of galaxies, but to further separate this cold component between the atomic and molecular phases one of many possible ``sub grid" recipes must be applied. These prescriptions are both theoretical and empirical in nature, and depend on quantities such as metallicity, dust-to-gas ratio, radiation field strength, and pressure \citep[e.g.][]{blitz06,krumholz09,gk11,krumholz13}.  

For the EAGLE simulation, \citet{lagos15} implement the recipes of \citet{gk11} and \citet{krumholz13}; both have a dependence on the radiation field and the amount of metals in the ISM, but the latter has an additional dependence on the atomic gas column density. Comparison with the accurate xCOLD GASS data points shows that neither recipe is particularly successful in predicting the correct amount of molecular gas in galaxies with $\log M_{\ast}/M_{\odot}<10.5$, with the \citet{krumholz13} prescription marginally preferred. MUFASA computes the molecular gas mass fractions on the fly using a recipe based on \citet{krumholz09}, which results in good agreement with HI gas fractions at the low-mass end and H$_2$ gas mass fractions for more massive galaxies, but over-predicts the molecular gas content of low-mass galaxies.  This in part arises owing to MUFASA's prescription that adds a subgrid clumping factor to \citet{krumholz09} to aid the resolution convergence of star formation rates, which for relatively low-resolution simulations such as MUFASA ends up yielding an increased molecular gas content.  As with the EAGLE comparison using two different H$_2$ formation recipes, this illustrates that detailed assumptions regarding the subgrid ISM model can significantly alter the predicted H$_2$ and HI content of simulated galaxies.  Hence while the gas content is emerging as a powerful constraint on models, which physical processes it actually constrains remains unclear owing to the significant differences in predictions between subgrid models even when using the same underlying simulation. A feature shared by all the hydrodynamic simulations investigated in Fig. \ref{modcomp} is that they robustly predict the atomic gas mass fraction in galaxies with $\log M_{\ast}/M_{\odot}>10.5$ but do not produce enough cold atomic gas in the most massive galaxies. This result highlights how the predicted HI gas fractions are far less dependent on the specific subgrid atomic-to-molecular conversion prescription applied, but that massive gas-rich galaxies are not successfully produced. 

Similarly, semi-analytic models are consistently reproducing very well the HI contents of galaxies with $\log M_{\ast}/M_{\odot}<10$. The one simulation shown in Fig. \ref{modcomp} that does the best at reproducing both molecular and atomic gas fraction scaling relations over the entire stellar mass range probed by xCOLD GASS is that of \citet{popping15}. This empirical model however also stands apart as being the only one which has used the molecular and atomic gas scaling relations as a constraint.  Of particular note is the overestimation of the molecular gas fractions in all the other semi-analytic models presented in Fig. \ref{modcomp}. As star formation is linked with the molecular phase, the fact that these simulations reproduce well the $z=0$ stellar mass function but not the \fgas\ scaling relation suggests that their detailed star-formation histories may not be accurate.  Whether they are used as a test of the output of simulations \citep[e.g.][]{lagos15,lacey16,dave17}, or as fitting constraints \citep[e.g.][]{popping15,martindale17}, the cold gas scaling relations, which have been observed and calibrated up to $z\sim2$ \citep{tacconi17}, offer significant opportunities to refine galaxy formation models and in particular to constrain subgrid recipes which are currently a significant source of uncertainty.

\section{Conclusions}

The important role of the gas cycle in the galaxy evolution framework is increasingly obvious, through observations of line and continuum emission in high redshift galaxies, as well as modelling and simulation work. The range of scales involved in this gas cycle are enormous: from $>$Mpc scales for the large but diffuse and hot gas reservoirs in the intergalactic medium, down through the circumgalactic and interstellar media, and to scales smaller than a parsec where star formation is taking place. 

We presented here the full and final data release of xCOLD GASS, the combination of two IRAM-30m large programmes: the original COLD GASS \citep{COLDGASS1}, which targeted galaxies with \mstar$>10^{10}$\msun, and its extension COLD GASS-low adding galaxies in the mass range $10^9-10^{10}$\msun. The xCOLD GASS ``survey philosophy" can be summarised by the following key points: 
\begin{enumerate}[I.]
\item Unbiased sample selection: to derive accurate and representative scaling relations between gas and global galaxy properties, a representative and unbiased sample is of the utmost importance. xCOLD GASS stands apart from previous large CO surveys \citep[e.g.][]{sanders85,kenney89,braine93,sage93,young95,boselli97,young11,lisenfeld11}, as it doesn't target galaxies in specific environments, or with particular morphologies or rates of star formation. The sample selection based only on stellar mass and redshift out of the SDSS DR7 spectroscopic sample makes possible the derivation of robust and representative scaling relations. 
\item Large sample size with dynamic range: the sample size of 532 galaxies is key as with it we can define both mean scaling relations and the scatter about the mean, as well as examine multi-parameter dependences. Furthermore, the flat stellar mass distribution (see Fig. \ref{histo}) ensures a large dynamic range for all the physical properties under consideration (e.g. \mstar, SFR, metallicity) while preserving the statistical relevance of the sample. 
\item Homogeneous and accurate measurements: all the physical properties provided as part of the survey have been measured across the sample using the same quality data products, and the same methods. This was done both for the molecular gas measurements themselves, all derived from consistent IRAM-30m observations, and the quantities measured from the ancillary multi-wavelength datasets (SDSS, WISE, GALEX,...)
\item Consistent observing strategy: by using a rigorous observing strategy, we can provide some meaningful and constraining upper limits on CO line fluxes for all the galaxies with line emission is not detected. This is essential to derive scaling relations for the entire galaxy population. 
\end{enumerate}

By adding the COLD GASS-low component to the original survey, thus forming xCOLD GASS, we have been able to extend our systematic investigation of the molecular gas contents of the local galaxy population to systems with stellar masses as low as $10^9$\msun\ with the following key results: 
\begin{enumerate}
\item The molecular gas mass fraction, \fgas$=M_{H_2}/M_{\ast}$, is on average 10\% in galaxies with \mstar$<5\times10^{10}$ \msun, and then plummets rapidly to 1\% as \mstar\ increases to $10^{11}$\msun. There is a direct link between this scaling of \fgas\ with \mstar\ and the shape of the main sequence in the SFR-\mstar\ plane. 
\item The strongest dependence of \fgas\ is on quantities tracking star formation, in particular SSFR where the correlation coefficient is $r=0.82$. While strong ($r=-0.63$) the correlation is weaker with \nuvr\ colour; unlike SSFR, this quantity is not corrected for dust extinction showing how molecular gas and dust-obscured star formation are strongly linked. 
\item By using the CO-to-H$_2$ conversion function of \citet{accurso17b} which is calibrated independently of any assumption on the relation between molecular gas and SFR, we are able to show that while star formation efficiency varies systematically across the galaxy population, stellar mass is not the key parameter driving these variations.
\item Through complementary observations of the CO(2-1) line with the APEX telescope, we derive a value of $r_{21}=0.79\pm0.03$ for the (2-1)/(1-0) CO luminosity ratio. This value is applicable for observations which as in the case of xCOLD GASS integrate over the entire ISM of galaxies, and is intermediate between values typically measured in the central and outer regions of nearby star-forming galaxies.  
\item The atomic and molecular gas mass fraction scaling relations have strong constraining power for galaxy formation models. We show that modern hydrodynamic and semi-analytic simulations which all reproduce well the $z=0$ stellar mass function, do not succeed as well in reproducing the cold gas scaling relations, unless those gas relations are used as constraints. 
\end{enumerate}

\acknowledgements
All the members of the xCOLD GASS team wish to warmly thank the staff at the IRAM observatory for the continuous help and support throughout the six years over which observations for the two large programmes were conducted. This study is also based on observations collected at the European Southern Observatory, Chile (proposal number 091.B-0593). 

We also thank Claudia Lagos and Gerg\"o Popping for making the results of their simulations available, and for discussing their specifics, and thank Taro Schimizu and Mike Koss for discussions on the question of molecular gas in AGN hosts. 

AS acknowledges support from the Royal Society through the award of a University Research Fellowship. BC, LC and SJ acknowledge support from the Australian Research CouncilÕs Discovery Project funding scheme (DP150101734). BC is the recipient of an Australian Research Council Future Fellowship (FT120100660).

\bibliographystyle{apj}
\bibliography{refs_amelie}

\begin{turnpage}
 
\begin{deluxetable*}{lcccccccccccccc}
\tablecaption{IRAM CO(1-0) and CO(2-1) measurements \label{COtab}}
\tablehead{
\colhead{ID} & \colhead{$\sigma_{CO10}$} & \colhead{FlagCO10} & \colhead{$S/N_{CO10}$} &   \colhead{$L^{\prime}_{CO10,cor}$} &  \colhead{$z_{CO10}$} & \colhead{$W50_{CO10}$} &  
\colhead{$\sigma_{CO21}$} & \colhead{FlagCO21} & \colhead{$S/N_{CO21}$} &  \colhead{$L^{\prime}_{CO21,obs}$} & \colhead{$z_{CO21}$} &  \colhead{$\alpha_{CO}$ \tablenotemark{a}} & \colhead{$\log M_{H2}$} \\ 
 & \colhead{[mK]} & &  &   \colhead{[$10^8$ K km/s pc$^2$]} & &  \colhead{$[\rm{km/s}]$} 
 & \colhead{[mK]} & & &   \colhead{[$10^8$ K km/s pc$^2$]} & &   & \colhead{$[\log M_{\odot}]$}  }
\startdata
124028 &   0.94 & 2   & \nodata &      0.22   & \nodata & \nodata &  1.06 & 1 &   4.85 &    $  0.08 \pm  0.02$ & 0.01769 &  9.17 &  8.30 \\
124012 &   2.22 & 1 &   7.46 &    $  1.12 \pm  0.24$ & 0.01780 & 182.9 &                                   \nodata &  \nodata & \nodata & \nodata & \nodata & 5.21 & $ 8.77 \pm 0.18$ \\
 11956 &   0.95 & 2   & \nodata &    1.05   & \nodata & \nodata &   1.41 & 1 &   5.68 &     $  0.81 \pm  0.22$ & 0.03951 &  3.67 &  8.59 \\
 12025 &   1.06 & 2   & \nodata &      1.12   & \nodata & \nodata & 1.24 & 2   & \nodata &  0.35   & \nodata &  2.86 &  8.51 \\
124006 &   1.74 & 1 &   8.44 &   $  0.85 \pm  0.18$ & 0.01779 & 173.2 &  1.64 & 1 &  14.13 &    $  0.37 \pm  0.08$ & 0.01777 &   5.56 & $ 8.68 \pm 0.18$ \\
124000 &   2.16 & 2   & \nodata &     0.50                       & \nodata  & \nodata & \nodata & \nodata & \nodata & \nodata & \nodata  & 6.31 &  8.50 \\
124010 &   3.32 & 1 &  16.23 &    $  4.51 \pm  0.81$ & 0.01774 & 112.7 &                                 \nodata & \nodata & \nodata & \nodata & \nodata & 3.15 & $ 9.15 \pm 0.17$ \\
 12002 &   1.18 & 2   & \nodata &     1.11   & \nodata & \nodata & 1.46 & 2   & \nodata &    0.42   & \nodata &  3.20 &  8.55 \\
124004 &   1.95 & 1 &   5.11 &   $  0.42 \pm  0.11$ & 0.01782 &  65.6 &    2.43 & 2   & \nodata &    0.05   & \nodata &  3.84 & $ 8.21 \pm 0.19$ \\
 11989 &   1.07 & 2   & \nodata &      1.20   & \nodata & \nodata &   2.14 & 2   & \nodata &   0.80   & \nodata & 2.86 &  8.54 \\
 11994 &   1.55 & 1 &   4.77 &     $  2.73 \pm  0.74$ & 0.03721 & 276.9 &     3.08 & 2   & \nodata &   0.87   & \nodata &  4.00 & $ 9.04 \pm 0.19$ \\
124003 &   0.87 & 2   & \nodata &     0.09   & \nodata & \nodata &    1.07 & 1 &   6.61 &     $  0.05 \pm  0.01$ & 0.01380 &    9.44 &  7.91 \\
 27167 &   1.17 & 2   & \nodata &     1.05   & \nodata & \nodata &  2.04 & 2   & \nodata &   0.57   & \nodata & \ 3.15 &  8.52 \\
  3189 &   1.24 & 1 &   5.99 &   $  2.69 \pm  0.64$ & 0.03833 & 238.6 &     2.40 & 1 &   3.22 &    $  0.67 \pm  0.25$ & 0.03840 &   4.53 & $ 9.09 \pm 0.18$ \\
124027 &   2.13 & 1 &   4.43 &   $  0.93 \pm  0.26$ & 0.01820 & 220.0 &     4.03 & 1 &   5.15 &    $  0.27 \pm  0.08$ & 0.01829 &   4.10 & $ 8.58 \pm 0.20$ \\
  3261 &   1.96 & 1 &   8.57 &    $  3.06 \pm  0.63$ & 0.03746 &  83.3 &    2.36 & 1 &  11.60 &   $  1.56 \pm  0.34$ & 0.03747 &  3.83 & $ 9.07 \pm 0.18$ \\
  3318 &   1.03 & 2   & \nodata &      1.19   & \nodata & \nodata &   1.89 & 2   & \nodata &  0.63   & \nodata &  2.95 &  8.55 \\
  3439 &   0.89 & 2   & \nodata &       0.97   & \nodata & \nodata &    1.44 & 2   & \nodata &   0.24   & \nodata &  3.28 &  8.50 \\
  3465 &   1.17 & 1 &   4.19 &   $  1.24 \pm  0.36$ & 0.02898 & 445.8 &    1.72 & 2   & \nodata &   0.38   & \nodata &  3.51 & $ 8.64 \pm 0.20$ \\
  \enddata
\tablenotetext{a}{Recommended CO-to-H$_2$ conversion factor calculated using the function of \citet{accurso17b}, in units of $M_{\odot}$(K \kms pc$^2$)$^{-1}$}
\tablecomments{The full version of this table, including all columns and all 532 xCOLD GASS galaxies, is available online. A detailed description of this table's contents, and the additional columns available online, is given in Section \ref{iramcat}.}
\end{deluxetable*}


 \end{turnpage}

\appendix
\section{A. xCOLD GASS spectral atlas}
\label{atlas}

\begin{figure}
\epsscale{1.0}
\plotone{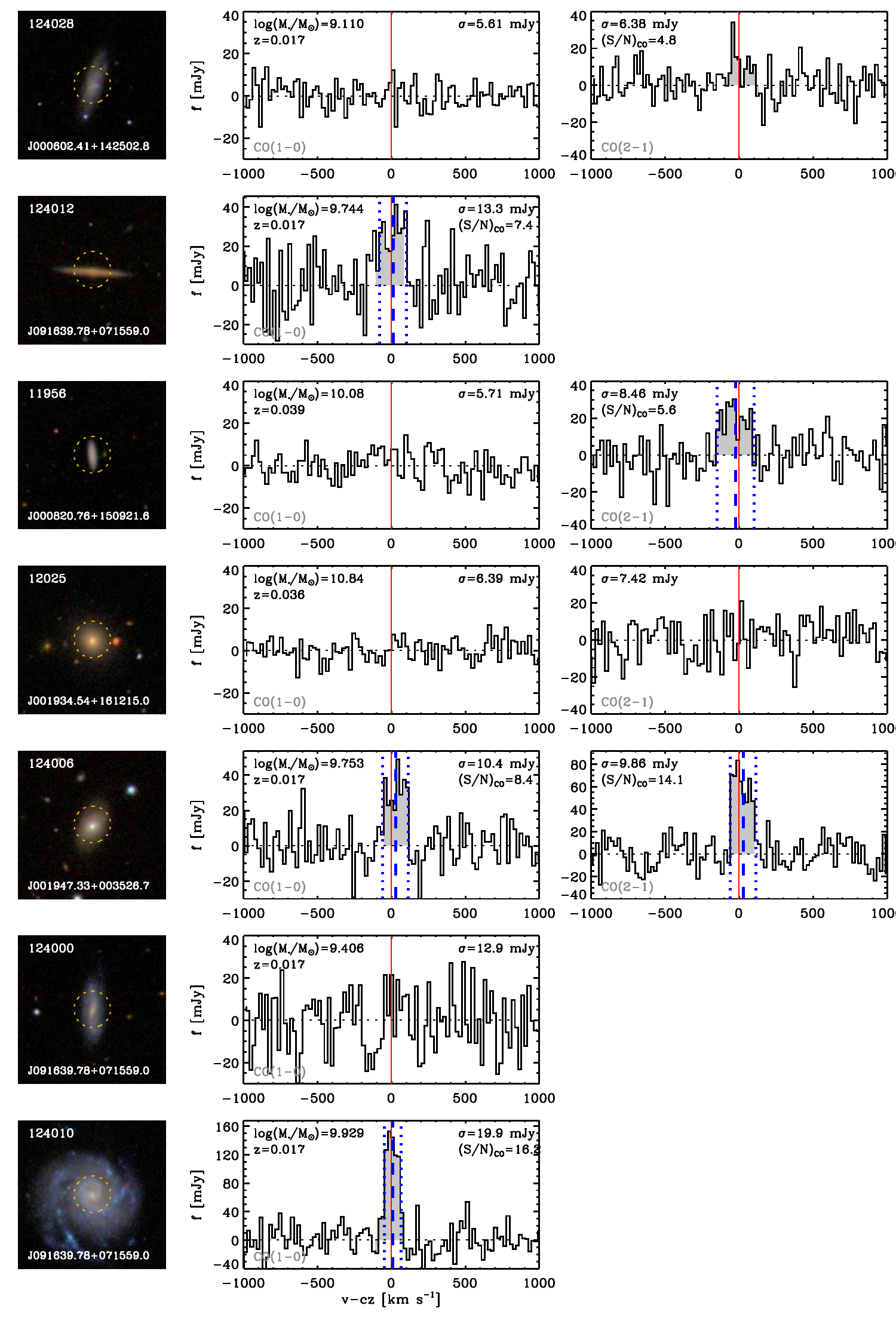}
\caption{SDSS images and IRAM-30m spectra of the xCOLD GASS galaxies. Each SDSS image is 1.5\arcmin$\times$1.5\arcmin\ in dimension, and shows the position and size of the IRAM-30m beam at the frequency of the CO(1-0) line (the diameter of the beam is half of that at the frequency of the CO(2-1) line).  The middle panel shown IRAM spectrum centered on the position of the CO(1-0) line. The solid red line is the expected center of the line based on the SDSS spectroscopic redshift. When the CO line is detected, the dashed blue line indicates the central velocity of the line and the dotted lines represent the FWHM of the line based on the fitting technique described in Sec. \ref{iramcat}.  {\edit The grey shaded area represents the region of the spectrum over which we integrated to calculate the total line flux.} The right hand panel shows the IRAM spectrum centered on the CO(2-1) line when those observations are available. \label{spec10}}
\end{figure}

\begin{figure}
\epsscale{1.0}
\plotone{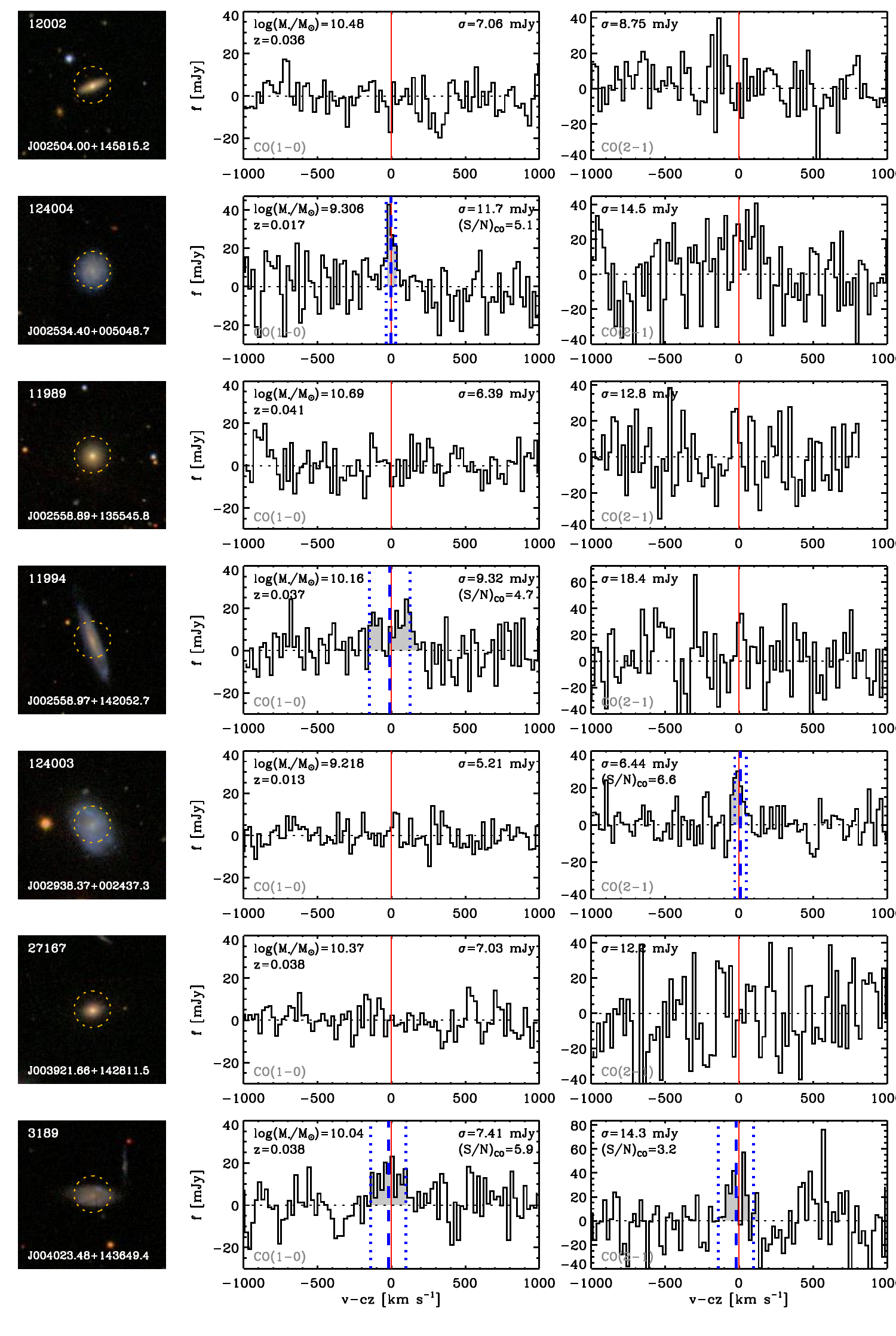}
\caption{SDSS images and CO(1-0) spectra.   \label{spec10}}
\end{figure}

\begin{figure}
\epsscale{1.0}
\plotone{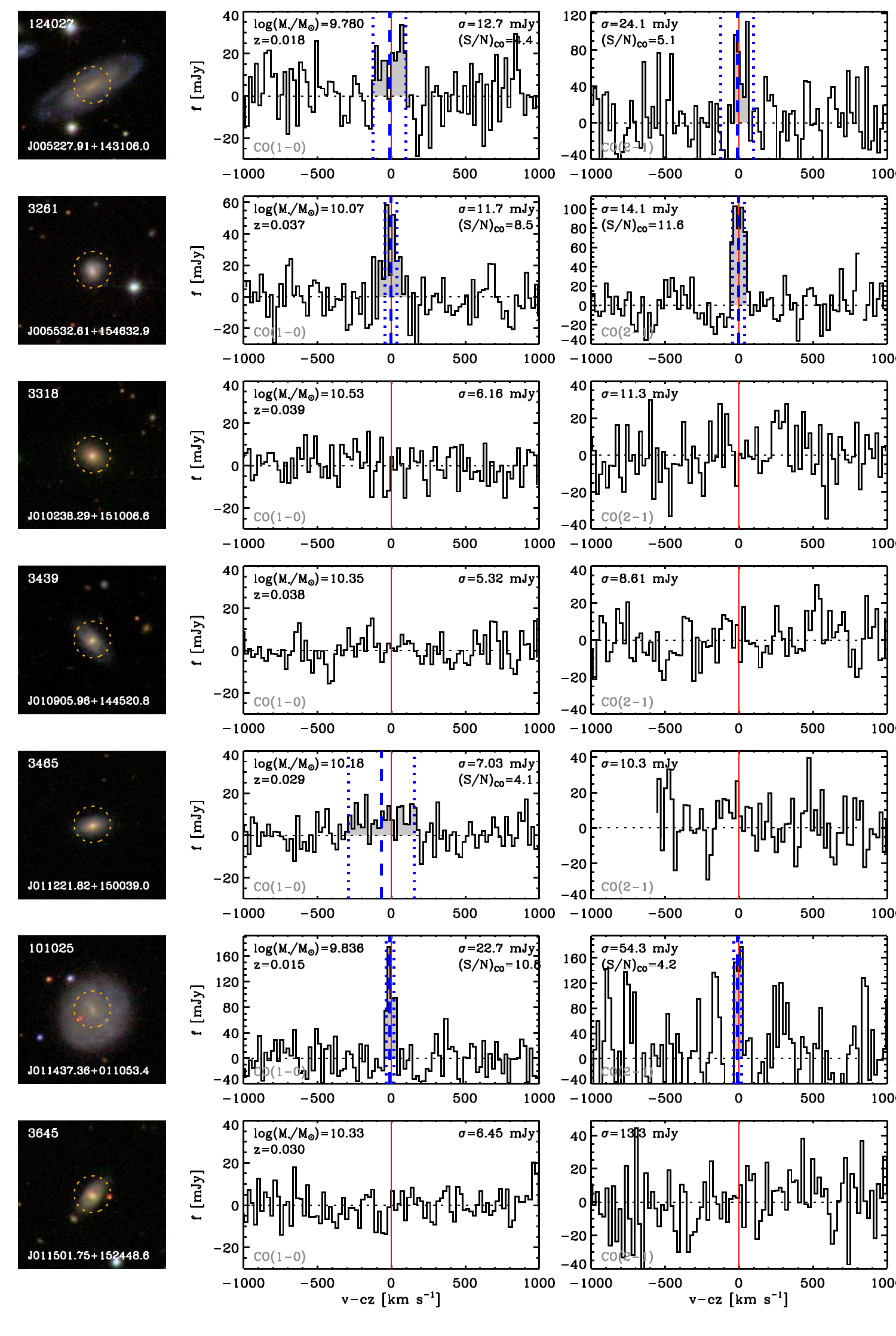}
\caption{continued from Fig. \ref{spec10}}
\end{figure}

\begin{figure}
\epsscale{1.0}
\plotone{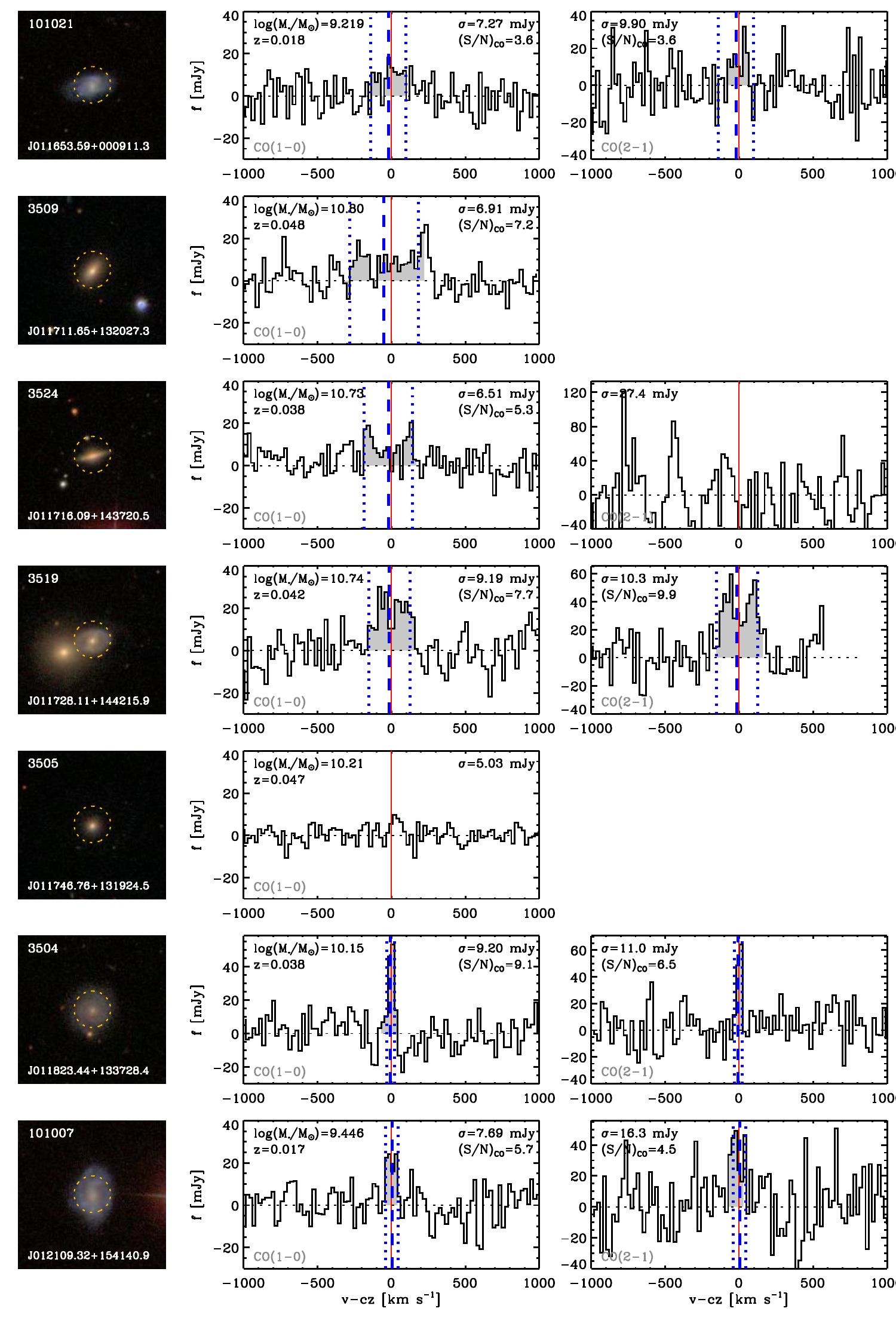}
\caption{continued from Fig. \ref{spec10} - the additional 73 figures showing the entire sample of 532 galaxies are available online.}
\end{figure}

\end{document}